\documentclass[useAMS]{mn2e}

\usepackage{amssymb,amsmath,graphicx,times}
\def\ltsima{$\; \buildrel < \over \sim \;$}
\def\simlt{\lower.5ex\hbox{\ltsima}}
\def\gtsima{$\; \buildrel > \over \sim \;$}
\def\simgt{\lower.5ex\hbox{\gtsima}}

\def\cm2{{cm$^{-2}$}}

\def\lum{{$L_{2-10}$}}

\def\p1{{Paper I}}

\def\xmm{{\em XMM--Newton}}
\def\chandra~{{\em Chandra}}

\def\chandra{{\em Chandra}}

\def\xmm{{\em XMM--Newton}}

\def\nh{{N$_{\rm H}$}}

\def\f14{{10$^{-14}$}}
\def\f13{{10$^{-13}$}}
\def\f12{{10$^{-12}$}}
\def\f11{{10$^{-11}$}}
\def\e22{{10$^{22}$}}

\def\3c{{3C 234}}
\def\l58{{$L_{5.8 \mu m}$}}

\title[The \chandra-COSMOS survey IV: X-ray spectra of the bright sample]{The \chandra-COSMOS survey IV: X-ray spectra of the bright sample}
\author[G. Lanzuisi et al.]
{G. Lanzuisi$^{1,2,3}$, F. Civano$^2$,  M. Elvis$^2$, M. Salvato$^1$, G. Hasinger$^4$, C. Vignali$^{5,6}$, 
\newauthor{G. Zamorani$^6$, T. Aldcroft$^2$, M. Brusa$^{5,6,1}$, A. Comastri$^6$, F. Fiore$^7$,}
\newauthor{A. Fruscione$^2$, R. Gilli$^6$, L. C. Ho$^8$, V. Mainieri$^9$, A. Merloni$^1$, A. Siemiginowska$^2$}
\\
\\
$^1$Max-Planck-Institut f\"ur extraterrestrische Physik,  Giessenbachstrasse 85748 Garching, Germany\\
$^2$Harvard-Smithsonian Center for Astrophysics, 60 Garden St. Cambridge, MA 02138.\\
$^3$Technische Universit\"at M\"unchen, Fakult\"at f\"ur Physik, James-Franck-Str. 1 85748 Garching, Germany\\
$^4$Institute for Astronomy, 2680 Woodlawn Drive Honolulu, HI 96822-1839 USA\\
$^5$Dipartimento di Astronomia, Universit\`a degli Studi di Bologna, Via Ranzani 1, I--40127 Bologna, Italy \\
$^6$INAF--Osservatorio Astronomico di Bologna, Via Ranzani 1, I--40127 Bologna, Italy\\
$^7$Osservatorio Astronomico di Roma (INAF), Via Frascati 33, I--00040  Monteporzio Catone, Italy \\
$^8$The Observatories of the Carnegie Institute for Science, Santa Barbara Street, Pasadena, CA 91101, USA\\
$^9$ESO Headquarters Karl-Schwarzschild-Str. 2 85748 Garching bei M\"nchen Germany\\
}

\begin{document}

\date{Submitted 2012 October 22}

\maketitle

\begin{abstract}

We present the X-ray spectral analysis of the 390 brightest extragalactic sources in
the \chandra-COSMOS catalog, showing at least 70 net counts in the 0.5-7 keV band.
This sample has a 100\% completeness in optical-IR identification, with $\sim$75\% of the sample having a 
spectroscopic redshift and $\sim$25\% a photometric redshift.
Our analysis allows us to accurately determine the intrinsic absorption, the broad band continuum shape ($\Gamma$) and intrinsic 
\lum\ distributions, with an accuracy better than $30\%$ on the spectral parameters for 95\% of the sample.
The sample is equally divided in type-1 (49.7\%) and type-2 AGN (48.7\%) plus few passive galaxies at low z.
We found a significant difference in the distribution of $\Gamma$ of type-1 and type-2,
with small intrinsic dispersion, a weak correlation of $\Gamma$ with \lum\ and a large population 
(15\% of the sample) of high luminosity, highly obscured (QSO2) sources.
The distribution of the X-ray/Optical flux ratio (Log(F$_X$/F$_i$)) for type-1 is narrow ($0<X/O<1$), while type-2
are spread up to $X/O=2$. The X/O correlates well with the amount of X-ray obscuration.
Finally, a small sample of Compton thick candidates and peculiar sources is presented.
In the appendix we discuss the comparison between \chandra\ and \xmm\ spectra for 280 sources in common.
We found a small systematic difference, with \xmm\ spectra that tend to have softer power-laws and lower obscuration.

\end{abstract}

\begin{keywords}
Galaxies:~active -- Galaxies:~high-redshift --
  Galaxies:~nuclei -- X-ray:~galaxies -- X-ray:~survey
\end{keywords}
  
\section{Introduction}

There is now strong observational evidence that the accretion onto super massive black holes (SMBHs)
and the formation and evolution of their host galaxies are strongly coupled processes.
A narrow correlation between the mass of SMBHs and the properties of 
their host galaxies is observed (Kormendy \& Richstone 1995, Silk \& Rees 1998, Magorrian et al. 1998, Ferrarese \& Merritt 2000).
There is also evidence for a common cosmic ``downsizing" in both  star formation rate (SFR) and active galactic nuclei (AGN) activity
(Cowie et al. 1996; Hasinger et al. 2005; La Franca et al. 2005; Bongiorno et al. 2007; Silverman et al. 2009; Serjeant et al. 2010).
These evidences imply that the evolution of galactic bulges and central BHs should be 
tightly linked through some sort of physical mechanism or feedback
(but see also Jahnke et al. 2011 for a different interpretation).
For example, luminous AGN can heat the interstellar
matter through winds, shocks and ionizing radiation, inhibiting star-formation and accretion itself
(Croton et al. 2006; Li et al. 2007; Hopkins et al. 2008, Cattaneo et al. 2009). 
To understand this feedback mechanism, and hence the evolution of the BH-Galaxy relation, beyond the local
Universe (Merloni et al. 2010, Peng et al. 2010), we need to constrain the overall AGN census in detail and
to determine the distribution of AGN parameters in classes and their evolution in redshift. 

X-ray surveys are the most effective way of selecting AGN, given that they are less affected by
obscuration, up to Compton thick (CT)\footnote{If the X-ray obscuring matter has a column density equal to or larger than the inverse of the Thomson cross-section, 
\nh$>\sigma_T^{-1}\sim1.6\times10^{24} cm^{-2}$, the source is called Compton thick.} regime,
and contamination form the host galaxy, that strongly affect optical and IR bands (e.g. Donley et al. 2012).
This is especially true for all those classes of 'elusive' AGN that are hard to identify as such: X-ray Bright Optically Normal Galaxies 
(XBONGs, Comastri et al. 2002, Civano et al. 2007);
high-z, obscured AGN in highly starforming galaxies (Alexander et al. 2002, 2005, Daddi et al. 2007, Fiore et al. 2008, 2009, Lanzuisi et al. 2009);
or local Low Luminosity AGN (LLAGN, Ho et al. 1995).
However, the combination with multi-wavelength spectroscopic and photometric observations
is required in order to determine the redshift of the source and the host properties, 
key information to fully understand how SMBHs interact with their environment.

It is still not clear why, to a first approximation, 
the spectral energy distribution of all quasars are very similar, over a large range of wavelengths
and of physical parameters such as BH mass, accretion rate, luminosity
and over 13 Gyr of cosmic time. 
Whatever processes produce the features we observe, 
there must be some robust physics that does not depend strongly on any of these parameters. 

While high signal to noise (S/N) sources show a variety of complex spectral features - warm absorbers, soft excess, 
emission and absorption lines, a reflection component
(e.g. Risaliti \& Elvis 2004 for a review) - 
for low S/N sources the typical X-ray (0.5-10 keV) spectrum can be modeled 
as a power-law, modified by intrinsic absorption from the surrounding material (the dusty 'torus' or
the host galaxy itself). 
The X-ray spectroscopy of large samples of AGN, with available adequate multi-wavelength coverage, 
allows one to study the distribution of at least three fundamental parameters such as the column density \nh, 
the photon index $\Gamma$\footnote{defined as $N(E)\propto E^{-\Gamma}$, where N(E) is the number of photons
of energy E.} and the intrinsic X-ray luminosity.

Typically all AGN show very similar photon indexes of the primary power-law, with average value $\Gamma\sim1.9$ and
small dispersion, $\sigma_{\Gamma}\sim0.2$ (Nandra \& Pounds 1994; Piconcelli et al. 2005; Tozzi et al. 2006; Mainieri et al. 2007).
This emission is due to the inverse Compton scattering of the UV photons, emitted from the disk,
by the hot electrons in the corona surrounding the disk (Haardt \& Maraschi 1999; Siemiginowska 2007).
This process tends to produce similar slopes, independently of the expected large range 
of physical parameters of the central engine, such as BH mass, spin or accretion rate.
To build a large sample of sources for which these fundamental parameters are known 
with a good accuracy is not trivial. Even the search for dependencies of the photon index
from other observables, such as luminosity, redshift or accretion rate, 
has not yet produced definitive results (Page et al. 2005; Mateos et al. 2005; Kelly et al. 2007; Young et al. 2009; Sobolewska \& Papadakis 2009, Lusso et al. 2012).

X-rays are very effective in probing the distribution of obscuration around the central SMBH, at least in the Compton thin
regime, revealing a large fraction of obscured sources both in the local Universe (Risaliti et al. 1999) and at high redshift (Tozzi et al. 2006). 
This obscuration can be used as an indicator that the host galaxy is undergoing feedback processes (Hopkins et al. 2008), 
i.e. the emission from the AGN is interacting with gas and dust in the environment.
The fraction of obscured AGN is also depending on the X-ray luminosity
(Lawrence \& Elvis 1982; Ueda et al. 2003; La Franca et al. 2005, Hasinger et al. 2006, Della Ceca et al. 2008).
Such a trend has been observed also in the optical (Simpson 2005) and in the
near infrared (Maiolino et al. 2007) and may be linked to the AGN radiative power (coupled with the age):
the AGN emission is thought to be able to ionize and expel gas and dust from the nuclear regions, as suggested in the so-called 
’receding torus’ models (Lawrence 1991; Ballantyne et al. 2006), indicating a breakdown of the standard AGN unification
model (Antonucci 1993).
The possible evolution of the fraction of obscured sources with
redshift is still matter of debate (La Franca et al. 2005; Treister \& Urry 2006; Hasinger et al. 2006, Gilli et al. 2007, Della Ceca et al. 2008).

Finally, the X-ray luminosity is the most efficient way to probe the total 
luminosity due to accretion onto SMBH,
because it is less affected by obscuration, with respect to Optical-UV radiation from the accretion disk itself.
Together with the BH mass, it allows to derive the accretion rate of the system.
Also, the luminosity function is the key ingredient to determine the cosmic history of accretion and the duty cycle of AGN.

We took advantage of the superb dataset provided by the \chandra\ coverage of the COSMOS field (Elvis et al. 2009, Paper I hereafter),
to investigate the X-ray emission of a large population of quasars, spanning a wide range of luminosities and redshift,
with the aim of studying the distribution of the spectral parameters within the sample and in different optical classes.
The high number of sources available with optical/IR identification, especially at high redshift and high luminosities,
makes this sample unique among X-ray deep surveys.
We limited our analysis to a bright subsample, in order to test our procedure and obtain strong constraints on the spectral parameters.

In the following we present the X-ray spectral analysis of the brightest 390 extragalactic sources in
the \chandra-COSMOS catalog, with at least 70 net counts in the 0.5-7 keV band.
The detection procedure is described in Puccetti et al. (2009, Paper II).
This \chandra-COSMOS Bright Sample (CCBS hereafter) has 100\% completeness in optical-IR identification (Civano et al. 2012, Paper III).
The paper is organized as follows: 
Section 2 summarizes the properties of the COSMOS survey; 
Section 3 presents the data reduction and sample general properties;
Section 4 reports the X--ray fit procedure and results; 
in Section 5 the fit parameter distributions are discussed, while 
section 6 summarizes the results.
A standard $\Lambda$ cold dark matter cosmology with $H_0=70$ km s$^{-1}$ Mpc$^{-1}$, $\Omega_\Lambda=0.73$ and $\Omega_M=0.27$ is assumed throughout the paper.
Errors are given at 1$\sigma$ confidence level for one interesting parameter, unless otherwise specified. 

\section{The COSMOS survey}

The Cosmic Evolution Survey (COSMOS, Scoville et al. 2007a) is a deep and wide
extragalactic survey which covers a 2~deg$^2$ equatorial (10$^h$, +02$^{\circ}$) field with imaging by most of the major space-based telescopes 
({\em Hershel}, {\em Spitzer}, {\em Hubble Space Telescope},  GALEX, \xmm\ and \chandra) 
and a number of large ground based telescopes ({\em Subaru}, VLA, ESO-VLT, UKIRT, NOAO, CFHT, and others).
Large dedicated ground-based spectroscopy
programs in the optical with Magellan/IMACS (Trump et al. 2007), VLT/VIMOS
(Lilly et al. 2008, 2009), {\em Subaru}-FMOS and DEIMOS-{\em Keck} have been completed or are well underway.
The location of COSMOS will allow all major future facilities (JVLA, ALMA, and SKA) to target this region. 
Newly awarded follow ups with \chandra, {\it Nu-Star}, JVLA and {\it Spitzer} will rejuvenate the database in the next years.

This wealth of data has resulted in a 35-band photometric catalog of
$\sim$10$^6$ objects (Capak et al. 2007) resulting in
photo-z's for the galaxy population accurate to $\Delta$z/(1+z)$<$1\%
(Ilbert et al. 2009) and to $\Delta$z/(1+z)$\sim$1.5\% for the AGN population
(Salvato et al. 2009, 2011).

The {\em Chandra}-COSMOS survey (C-COSMOS) covered the
central 0.9~deg$^2$ region of the COSMOS field,
with the ACIS-I CCD imager (Garmire et al. 2003) on board the \chandra\ X-ray Observatory (Weisskopf et al. 2002). 
The survey used a total of 1.8~Msec of \chandra~observing time ($\sim$21~days), 
employing a series of 36 heavily overlapping ACIS-I 50~ksec
pointings to give a highly uniform exposure of $\sim$160~ksec over the inner (0.5 deg$^2$) area
and $\sim70-80$ks in the outer (0.4 deg$^2$) area.
The depth is $\sim$1.9$\times$10$^{-16}$erg~cm$^{-2}$s$^{-1}$ (0.5-2~keV), three times below the corresponding flux limits for
the XMM-COSMOS survey (Hasinger et al. 2007; Cappelluti et al. 2009), making them 
complementary surveys.
The X-ray source catalog comprises 1761 X-ray point sources detected down to a maximum
likelihood threshold $detml = 10.8$\footnote{This likelihood threshold corresponds to
a probability of $\sim5\times10^{-5}$ that a catalog source is a background fluctuation.}
in at least one band (0.5-2, 2-8, or 0.5-8 keV). 
The C-COSMOS survey, data analysis and general properties of the detected sample are described 
in detail in Paper I.
Paper II presented  the details of the simulations carried out to optimize the 
detection procedure.
The optical and infrared identifications of the C-COSMOS sources have been presented 
in Paper III.

In the latest \chandra\ call for proposal (Cycle 14), the extension of the coverage
of the COSMOS field has been approved (P.I. F. Civano).
The proposal aims to complete, with 2.8 Msec of observation, the deep Chandra coverage of the COSMOS area,
taking $\sim1.45$ deg$^2$ to 160 ks depth and covering the remaining 0.75 deg$^2$ field to 100-50 ks depth,
more than doubling the number of detected sources.

\section{The C-COSMOS Bright Sample}

\subsection{Counts extraction}

The extraction of source and background counts, and the creation 
of response matrices, ARF and RMF, for each source, has been performed 
with {\it Yaxx}\footnote{http://cxc.harvard.edu/contrib/yaxx/}. 
{\it Yaxx} is a {\it Perl} code, developed by T. Aldcroft at the Harvard 
Smithsonian Center for Astrophysics, which automatically deals with
the extraction and fitting of sources detected in large \chandra\ and \xmm\ surveys,
such as ChaMPs (Kim et al. 2004) and COSMOS. 
{\it Yaxx} makes use of standard \chandra\ CIAO\footnote{http://cxc.harvard.edu/ciao} (Fruscione et al. 2006)
and \xmm\ Scientific Analysis Software (SAS\footnote{http://xmm.esac.esa.int/sas/current/index.shtml})
tools for the spectra extraction and ARF and RMF creation.
CIAO v. 4.1.2 software and CALDB 4.1.2 calibration files were used for the extraction.

The extraction of counts of all the sources listed in an input catalog, 
is performed in each observation separately, and then the counts 
are merged together. Given the C-COSMOS tiling, any individual source
can be detected up to four times.
The radius of the source extraction regions take into 
account the degradation of the PSF with off-axis angle, 
thus it can change significantly from one 
observation to the other for the same source, in a range of 2\arcsec-10\arcsec. 
Each radius is chosen in order to optimize the signal to noise ratio (S/N) in each observation. 
The background extraction annuli have an inner radius slightly larger than the source radius, 
and a fixed outer radius of 30\arcsec.
In most cases, the extraction regions had to be masked to avoid CCD gaps, 
field of view (FOV) edges and other sources. 
These regions were excluded from the extraction areas.

The response matrices, ARF and RMF, are computed with standard CIAO tools in {\it Yaxx}.
The source and background spectra are then merged together using the
{\it mathpha} tool , and the total ARF and RMF are computed with {\it addarf} 
and  {\it addrmf} tools, within the FTOOLS package (Blackburn 1995).
The scaling factor of source and background regions changes in each observation, 
and this is taken into account when producing the final merged events. 
We developed a procedure in order to ensure that the scaling factor of the final 
merged event is equal to the mean of the scaling factors of the different observations, weighted for the 
exposure times (i.e. the contribution to the total number of counts, assuming constant flux).
This procedure was performed for all the 1761 sources in the catalog, 
and the resulting distribution of total net counts defined in this way
was used to select our sample.

\begin{figure*}
\begin{center}
\label{sample}
\includegraphics[width=8cm,height=8cm]{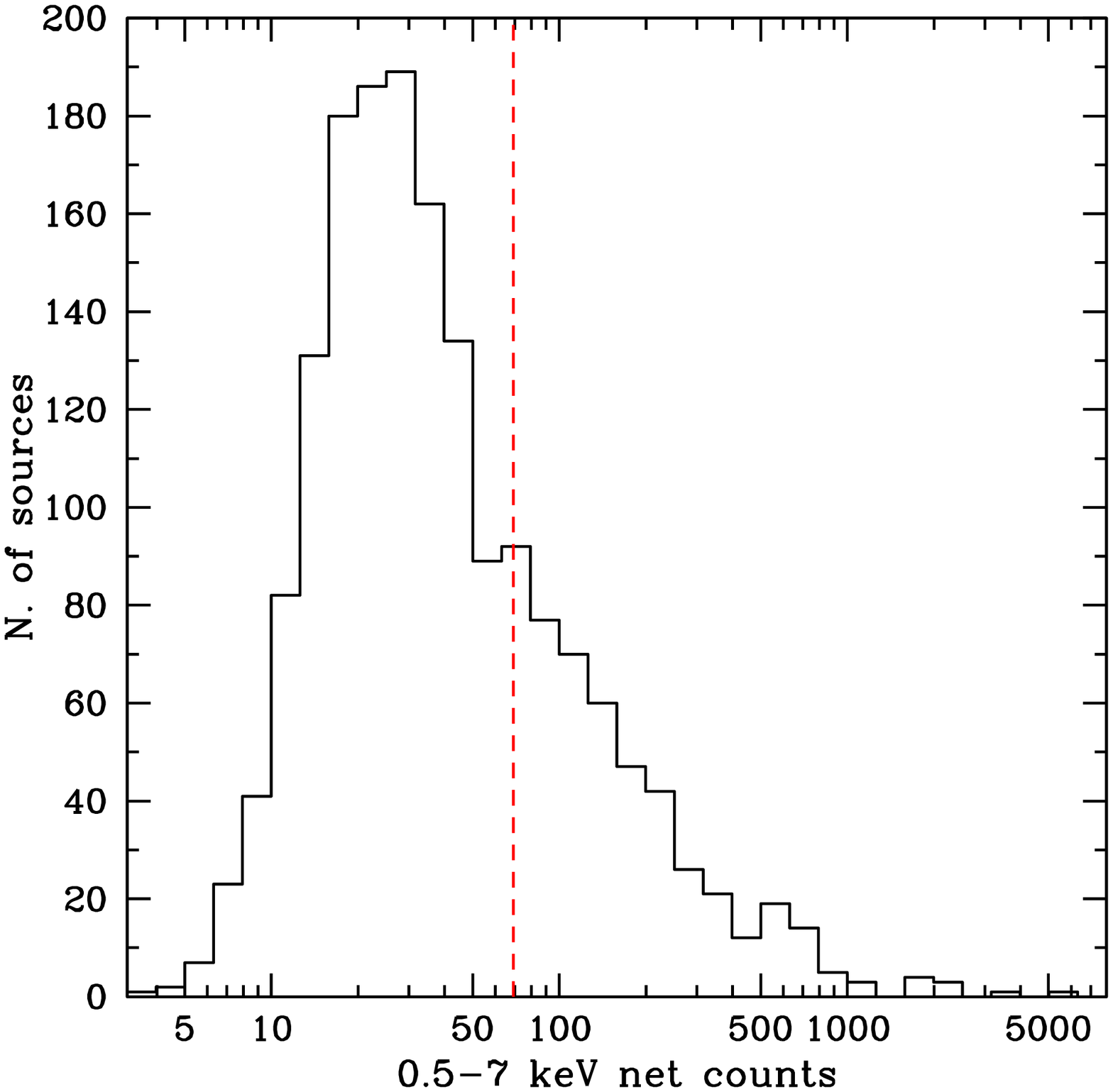}\hspace{1.5cm}\includegraphics[width=8cm,height=8cm]{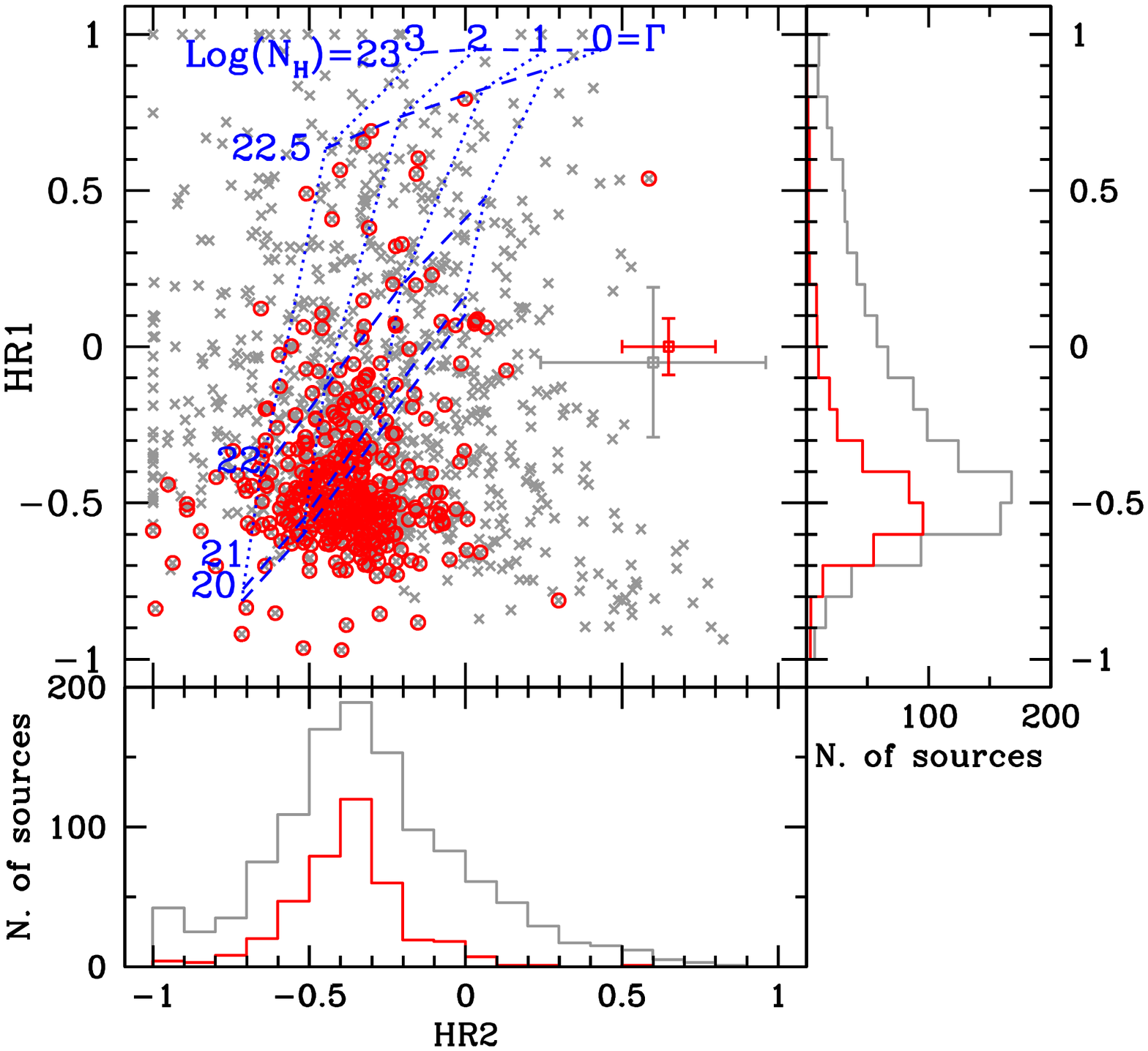}
\caption{{\it Left panel:} Distribution of 0.5-7 keV net counts of the total \chandra\ source catalog (1761 sources). 
The red dashed line marks the threshold we chose in the selection of the 405 sources in the CCBS (i.e. $>70$ counts). 
{\it Right panel:} X-ray color color diagram for the total \chandra\ catalog (grey points).
For clarity only the 1168 sources with errors on HR1 and HR2 smaller than 0.5 are plotted.
The grey cross represents the average error for this sub-sample.
The 405 sources in the CCBS are labeled with a red circle.
The red cross represents the average error for the CCBS.
Dashed and dotted lines show the expected HR for simple power-law spectra
with intrinsic absorption of log(\nh)= 20, 21, 22, 22.5, 23 cm$^{-2}$ (from bottom to top)
and photon index $\Gamma$= 3, 2, 1, 0 (from left to right), respectively. 
The right and lower panels show the distribution of HR1 and HR2 for the total sample (grey) and the CCBS (red).}
\end{center}
\label{istoc}
\end{figure*}

Among the 1761 sources in the C-COSMOS catalog, we selected sources with enough net counts to allow a spectral fit that could 
simultaneously determine the power law photon index $\Gamma$, normalization, and the intrinsic column density \nh\ of each source
with reasonable uncertainties and minimum correlation.
We found that selecting sources with more than 70 net counts in the 0.5-7 keV band and using the Cash 
statistic (Cash 1979), we were able to determine the spectral parameters with an 
accuracy of at least 30\% in most cases (see Sec. 4.2).
The initial resulting \chandra-COSMOS bright sample (CCBS) has 405 sources.

Fig. 1 (left panel) shows the distribution of 0.5-7 keV net counts for 
all the 1761 sources in the catalog, 
and the threshold of 70 counts chosen to select the CCBS (23\% of the total sample).

\subsection{X-ray colors}

In order to evaluate how the X-ray properties of the sources in the CCBS relate to
those of the entire C-COSMOS sample,
we computed two X-ray colors (hardness ratios, HR) for all the 1761 sources. 
The hardness ratios are defined as HR1=(M-S)/(M+S) and HR2=(H-M)/(H+M),
where S, M and H are the net source counts in the 0.5-2, 2-4 and 4-7 keV bands respectively.
Fig. 1 (right panel) shows the X-ray color-color diagram for the 
entire \chandra\ sample (grey points).
For clarity only sources with errors on HR1 and HR2 smaller than 0.5 are plotted (1168 sources out of 1761).
The 405 sources in the CCBS are marked with red circles.

Dashed and dotted lines show the expected HRs for simple power-law spectra
with intrinsic absorption (in the observer frame) of log(\nh)= 20, 21, 22, 22.5, 23 cm$^{-2}$ (from bottom to top)
and photon index $\Gamma$= 3, 2, 1, 0 (from left to right) respectively.
The right and lower projections show the distribution of HR1 and HR2 for the total sample (grey) and the CCBS (red).
The sources in the CCBS reproduce the distribution of X-ray colors of the total 
sample in the range of HR1 and HR2 $= -0.7-0$, where the 
majority of the sources lie, but CCBS is biased toward sources with softer spectra (negative HRs) with respect to the total sample.
The CCBS excludes most of the outliers,
those with HR1 and HR2 $>$ 0, 
i.e. sources with flat or inverted spectra.
Almost all the sources not detected in one of the bands (HR1 or HR2 $= \pm1$) are also excluded. 
The 11 sources with HR1 $< -0.7$ and HR2 $<-0.1$, i.e. sources with extremely soft spectra,
are all comprised in the CCBS and are optically identified as stars (see Sec. 3.3). 
Indeed the CCBS represents just the tip of the iceberg of the complete catalog and a large 
number of interesting sources such as heavily obscured, CT AGN 
and sources with peculiar spectra are located below the threshold of 70 net counts. 
Thus the extension of this study to fainter sources
is certainly of great interest, even if beyond the scopes of this paper.

\begin{figure}
\begin{center}
\includegraphics[width=8cm,height=8cm]{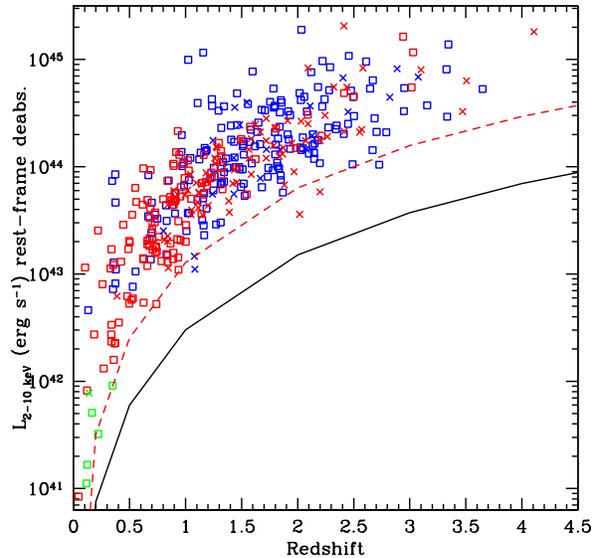}
\caption{Distribution of the 2-10 keV rest frame luminosity, corrected for absorption, as computed in Sec. 5, 
as a function of redshift for the 390 sources in the CCBS. 
In blue are plotted type-1 AGN, in red type-2 AGN and in green galaxies. 
Open squares (crosses) represent spectroscopic (photometric) redshifts. 
The solid line represents the sensitivity limit of the whole survey, while
the dashed line represent the flux limit estimated for the CCBS. 
}
\end{center}
\label{zlum}
\end{figure}

\subsection{Optical-IR identification}

The sources in the CCBS have a 99\% identification rate.
Eleven sources (2.7\% of the sample) are classified as stars (Wright, Drake and Civano 2010).
Four are not identified: a possible counterpart exists but is close to bright
stars or galaxies, for which reliable photometry is not possible
and there is not an entry in any COSMOS photometric catalog
(see Paper III for details).
We excluded these 15 sources from the sample, to focus on identified extragalactic 
sources, and we will refer in the following only to these 390 extragalactic sources as CCBS.
$\sim$75\% (295 sources) of the sample have a 
spectroscopic redshift and the remaining $\sim$25\% (95 sources) a photometric redshift 
from Salvato et al. (2009, 2011) or Ilbert et al. (2009).

The sample has been divided in different classes on the basis of the optical spectral 
information or the best fit template used in the SED fitting procedure, following 
the criteria presented in paper III and Brusa et al. (2010) for the identification. 
The sources are divided as follows:

\begin{table*}
\begin{center}
\caption{Distribution of CCBS sources in different classes, and their average redshift and X-ray properties.}
\begin{tabular}{ccccccccc}
\hline\hline\\
\multicolumn{1}{c} {Class} & 
\multicolumn{1}{c} {Num.}  &        
\multicolumn{1}{c} {\%} &          
\multicolumn{1}{c} {z$_{min}$, z$_{max}$} &
\multicolumn{1}{c} {\% z$_{spec}$}&
\multicolumn{1}{c} {$\langle z_{spec}\rangle$, $\langle z_{phot}\rangle$}& 
\multicolumn{1}{c} {$\langle N_{cts}\rangle$}&       
\multicolumn{1}{c} {$\langle F_{2-10 keV}\rangle$} &     
\multicolumn{1}{c} {$\langle L_{2-10 keV}\rangle$} \\   
 (1) & (2) &(3) & (4) & (5) & (6) & (7) & (8) & (9)\\  
\hline\\  
Type-1   &  194  &49.7 &  0.128-3.650 & 89.2 & 1.612,1.808  & 310.5 & $1.6\times10^{-14}$ &  $2.4\times10^{44}$ \\
Type-2   &  190  &48.7 &  0.045-4.108 & 61.6 & 0.919,1.666  & 166.3 & $1.4\times10^{-14}$ &  $1.4\times10^{44}$ \\
Galaxies &    6  & 1.5 &  0.116-0.350 & 83.3 & 0.194,0.141  & 91.2  & $5.2\times10^{-15}$ &  $4.6\times10^{41}$ \\
\hline                                                     
\hline                              
\end{tabular}
\end{center} 
(1) Source class; (2) number of sources; (3) fraction the CCBS sample in each class; (4) minimum and maximum
redshift; (5) fraction of spectroscopic redshift; (6) average of the spectroscopic and photometric redshifts; 
(7) average 0.5-7 keV net counts; (8) average 2-10 keV flux; (9) average 2-10 keV luminosity.
\end{table*}

\begin{itemize}
\item 194 (49.7\% of CCBS) are classified as type-1 AGN. 
Of these, 173 have a spectroscopic redshift showing at least one broad (FWHM$>2000$ km/s) emission line
and are therefore identified as Broad Line AGN (BLAGN). 
The remaining 21 have a photometric redshift, and the best fit templates are 
of unobscured quasars of various luminosity with different degrees of 
contamination by star-forming galaxy templates (from 10 to 90\% of the Optical-NIR flux).
\item 190 (48.7\% of CCBS) are classified as type-2 AGN.
Of these 117 have a spectroscopic redshift and have been identified as
non-broad line, i.e. a combination of Narrow Line AGN (NLAGN) and emission or absorption line galaxies. 
73 have a photometric redshift, and their best fit SED templates are a combination of obscured AGN templates 
with different degrees of contamination by passive galaxies, or pure spiral or star-forming galaxy templates.
All ``passive" galaxies with X-ray luminosity L$_{2-10 keV}>10^{42}$ erg s$^{-1}$
are classified as type 2 AGN (see below).
\item 6 (1.5\%) are classified as galaxies. Five have a spectroscopic redshift 
and an optical spectrum typical of passive or pure star-forming galaxies. One has a 
photometric redshift with a best fit template of passive galaxy.
All have  L$_{2-10 keV}<10^{42}$ erg s$^{-1}$.
\end{itemize}
Table 1 summarizes the distribution of the CCBS sources in different classes.
The minimum and maximum redshift, fraction of spectroscopic redshift, average spectroscopic and photometric redshift,
average 0.5-7 keV net counts and 2-10 keV flux and luminosity are reported.

Among the 117 sources classified as type 2 on the basis of their
optical spectrum and X-ray emission, 32 have a spectrum of NLAGN, according to their position in the diagnostic diagrams presented in  Bongiorno et al. (2010).
73 are classified as Emission line galaxies.
Of these, 69 have spectral coverage or spectral quality that do not allow to place 
the sources in the diagnostics diagrams, most of them being at z$>$1, where these diagrams cannot be constructed,
while only 4 are confirmed star forming or LINERs.
The remaining 12 are classified as absorption line galaxies (Trump et al. 2009).

Furthermore, among the 73 sources classified as type 2 on the basis of their SED fitting, only 11 have an 
SED template of type 1.8-2 Seyfert, 16 have a spiral, starforming or ULIRG template from Salvato et al. (2009),
while 46 have a template from the library of passive galaxies from Ilbert et al. (2009), 
(see flow chart in Fig. 6 of Salvato et al. 2011 for details on the template class assignation).

However, all these sources have an X-ray luminosity exceeding the value of L$_{2-10 keV}=10^{42}$ erg s$^{-1}$.
We used this value as a limit to discriminate between star formation and accretion processes.
Using the relation in Ranalli et al. (2003) this hard X-ray luminosity correspond to
a SFR$\sim200$ M$_{\odot}$/yr, in the ULIRG regime, i.e. they should be quite vigorous (and rare) starbursts.
We also stress that 92.5\% of the CCBS sample have L$_{2-10 keV}>10^{43}$ erg s$^{-1}$,
that would corresponds to SFR$\sim2000$ M$_{\odot}$/yr, if produced by star formation.
There are no sources with this SFR in the COSMOS field (Feruglio et al. 2010)
and therefore at the X-ray luminosities of the CCBS
we can be sure at least of the presence of an AGN, 
even if the presence of both components is not excluded.
We therefore apply an a-posteriori X-ray luminosity cut, and classified all these sources with
L$_{2-10 keV}>10^{42}$ erg s$^{-1}$ as type 2 AGN.

These results show that optical diagnostic diagrams and SED model fitting can be insensitive to hybrid objects (e.g. obscured AGNs
with enhanced star formation).
Indeed 70\% of the type-2 with an optical 
spectrum, and $\sim80\%$ of type-2 with a photo-z from SED fitting, would have been 
classified as starforming or passive galaxies (or would remain unclassified) on the basis of the optical/IR data only.
The coupling of diagnostic diagrams and SED fitting, with the X-ray luminosity cut
is indeed a more effective method of separating obscured AGNs from truly nonactive
galaxies (see also the discussion in Paper III and Trouille et al. 2011),
 especially at these flux levels and in the cases of hybrid objects or in all the cases in which the 
AGN contribution is overwhelmed in optical/IR by the host-galaxy light.

Fig. 2 shows the 2-10 keV rest frame luminosity, corrected for absorption 
as computed in Sec. 5, as a function of the redshift for the CCBS. 
Empty squares (crosses) represent sources with spectroscopic (photometric) redshift.
Blue, red and green colors represent type-1, type-2 and galaxies, respectively.
The sensitivity limit of the total \chandra-COSMOS survey is represented by the solid line, 
while the sensitivity limit
estimated for the flux of a source with 70 net counts and a typical AGN spectrum 
(a simple power law with $\Gamma=1.9$, \nh$=1\times10^{21}$ cm$^{-2}$) is reported with a dashed line.

\section{Spectral analysis}

\begin{figure*}
\begin{center}
\includegraphics[width=8cm,height=8cm]{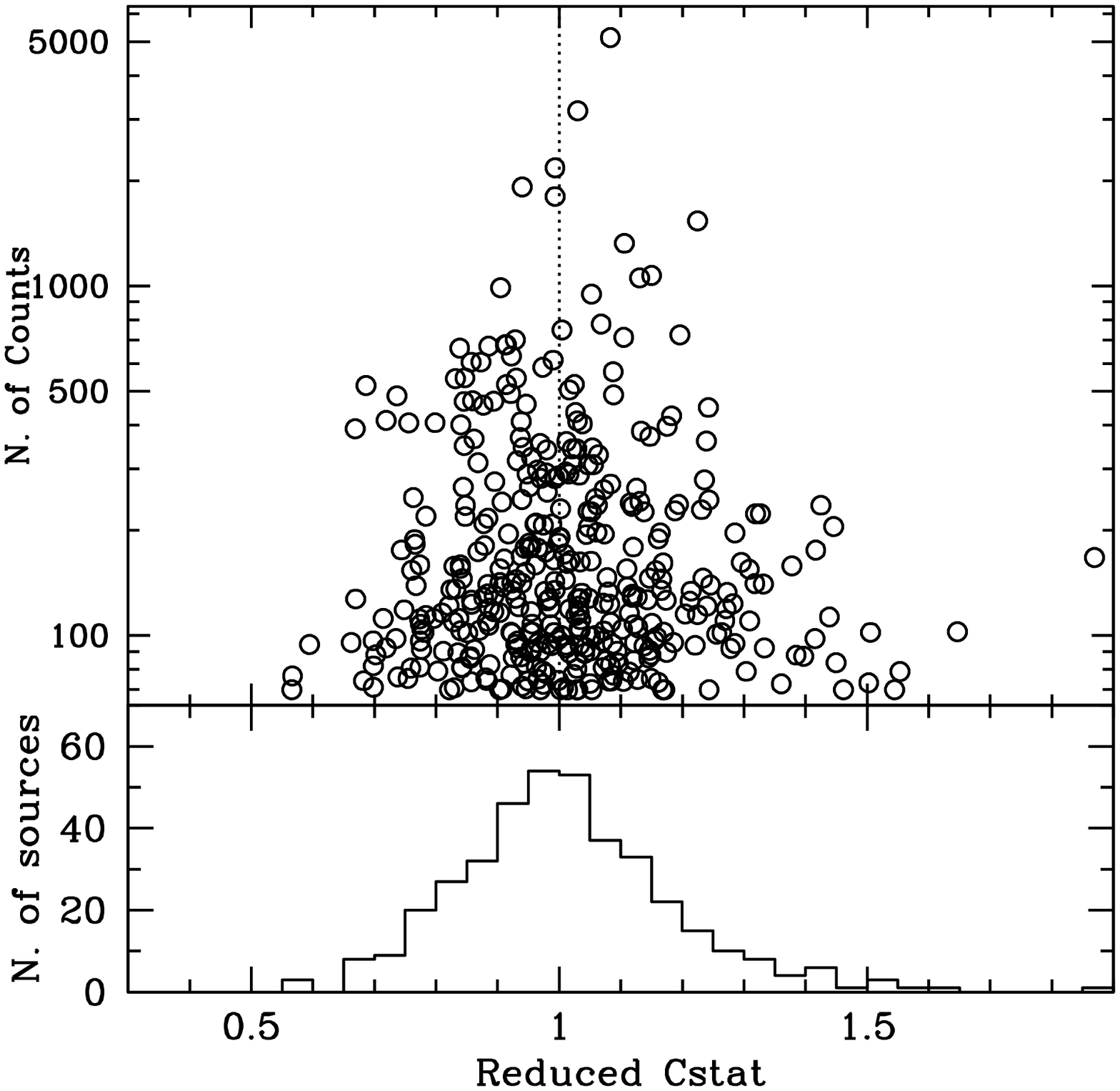}\hspace{1.5cm}\includegraphics[width=8cm,height=8cm]{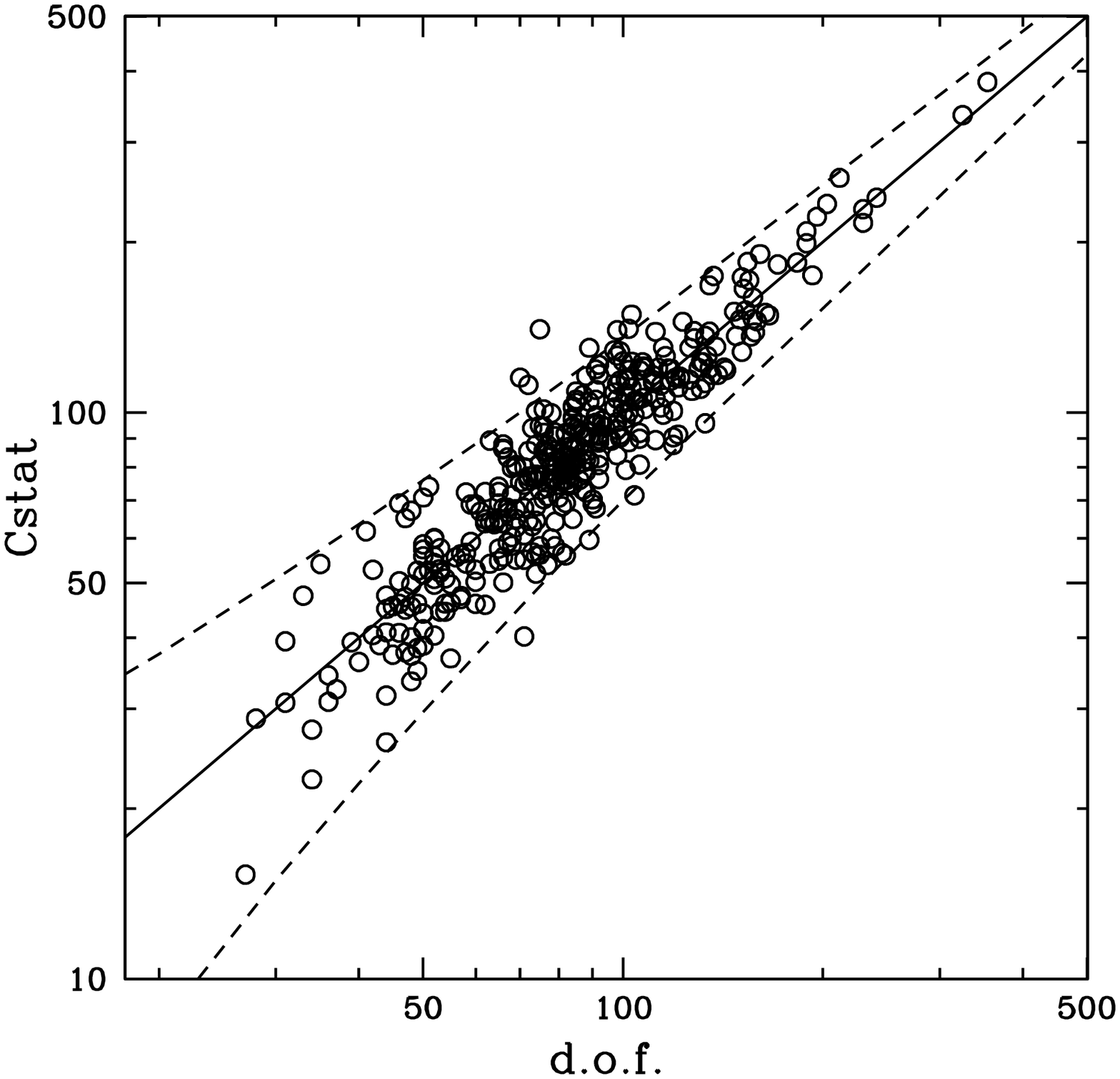}
\caption{{\it Left panel:} The reduced Cstat (Cstat$_\nu$) as a function of the 0.5-7 keV net counts for the CCBS sample is shown.
The bottom panel shows the histogram of the Cstat$_\nu$ for the sample. 
The dotted line shows the expected value for a good fit of Cstat$_\nu=1$.
{\it Right panel:} Distribution of Cstat with respect to DOF
for each source. The thick line represents the ideal line of Cstat$_\nu=1$,
while the dashed lines show, for a given number of DOF, the value of Cstat
above (below) which we expect a 1\% probability to find such 
high (low) value if the model is correct.}
\end{center}
\label{cstat}
\end{figure*}

\begin{figure}
\begin{center}
\includegraphics[width=8cm,height=8cm]{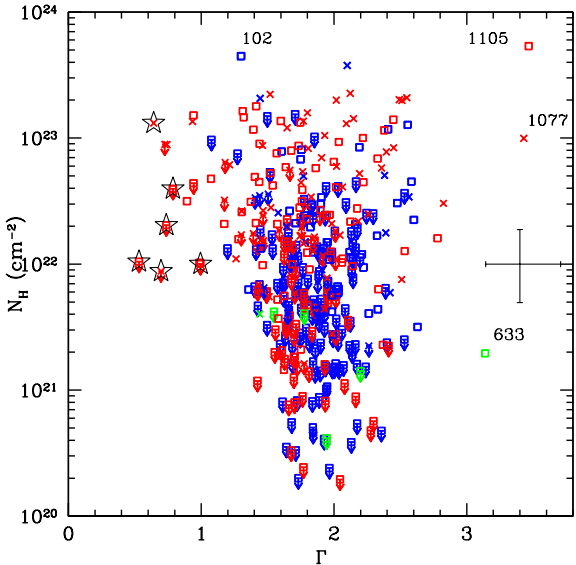}
\caption{Distribution of $\Gamma$ and \nh~for the CCBS sample.
Arrows represent 1$\sigma$ upper limits in the value of \nh.
Symbols as in Fig. 2. Peculiar sources discussed in Sec. 4.3 are labeled with the CID number.
Black stars are CT candidates on the basis of their photon index, discussed in Sec. 4.3.5.}
\end{center}
\label{gamnh}
\end{figure}

\subsection{Spectral fit}

The spectral fitting was performed using {\it Sherpa}\footnote{http://cxc.harvard.edu/sherpa/index.html}, 
the CIAO  modeling and fitting package (Freeman, Doe and Siemiginowska 2001).
We performed a spectral fit based on the Cstat statistics, a modified version
of the Cash statistic (Cash 1979), which is well suited for the low counts regime
because, in principle, it does not require counts binning.
The advantage of Cstat as implemented in the fitting package {\it Sherpa}, 
with respect to the Cash statistic, is that
the change in Cstat from one model fit to the next 
($\Delta C$), is distributed approximately as $\Delta\chi^2$, 
and it is possible to estimate the goodness of the fit\footnote{http://cxc.harvard.edu/sherpa/statistics/index.html},
i.e., the observed statistic, divided by the number of degrees of freedom (DOF), should be of order 1 for good fits.
However, this is true only when the number of counts in each bin is $>>1$.
Therefore, we binned the spectra to 5 counts per bin (see Appendix A for details).

Cstat is a maximum likelihood function and assumes that the error on the counts is purely Poissonian,
and so it cannot deal with data that have already been background subtracted.
Therefore, one has to model the background, and include this model in the fit.
In the {\it Sherpa} package this is accomplished in one step by performing a simultaneous fit of the background 
and the total (source+background) spectra.

\begin{figure*}
\begin{center}
\includegraphics[width=8cm,height=8cm]{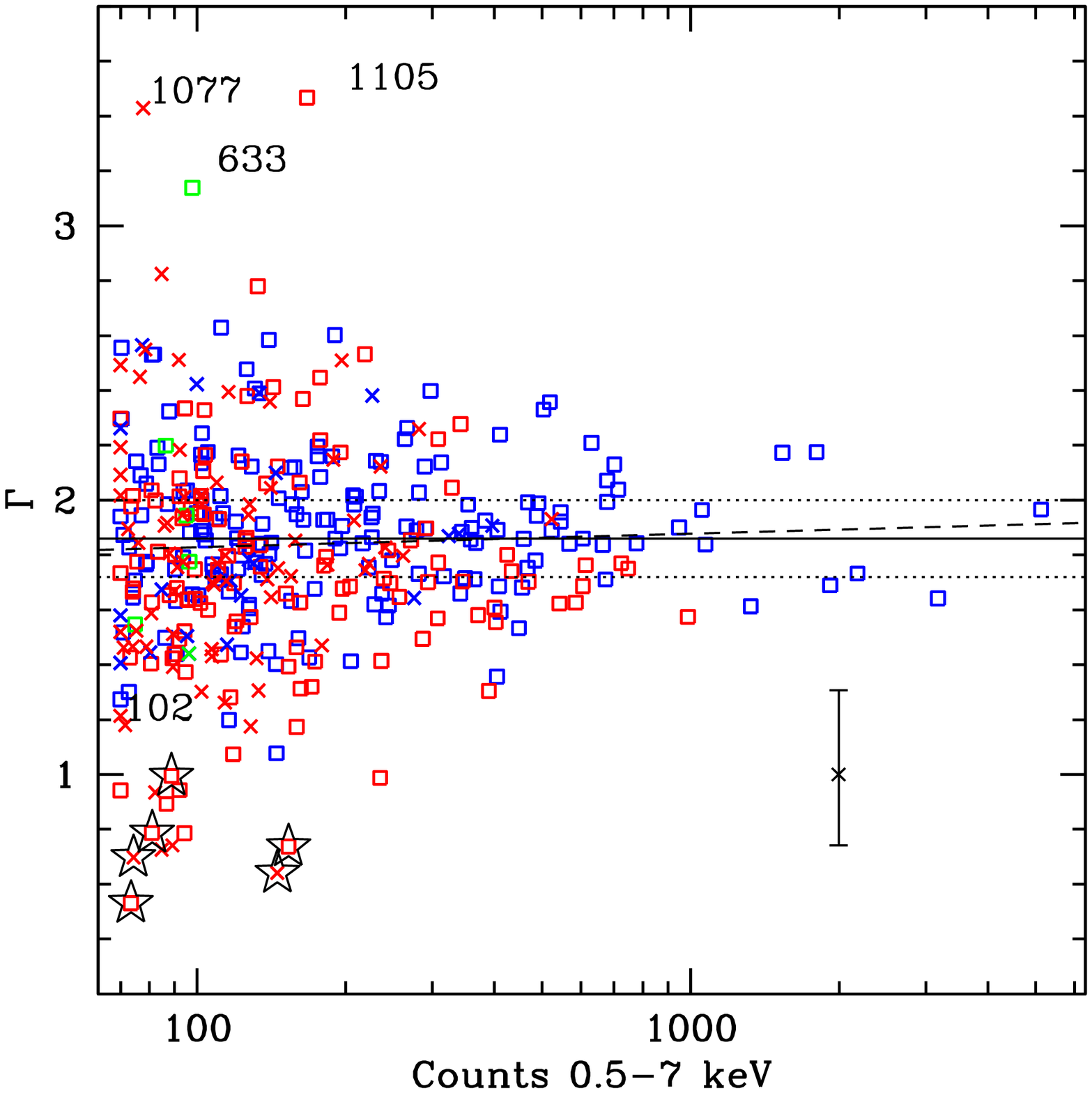}\hspace{1.5cm}\includegraphics[width=8cm,height=8cm]{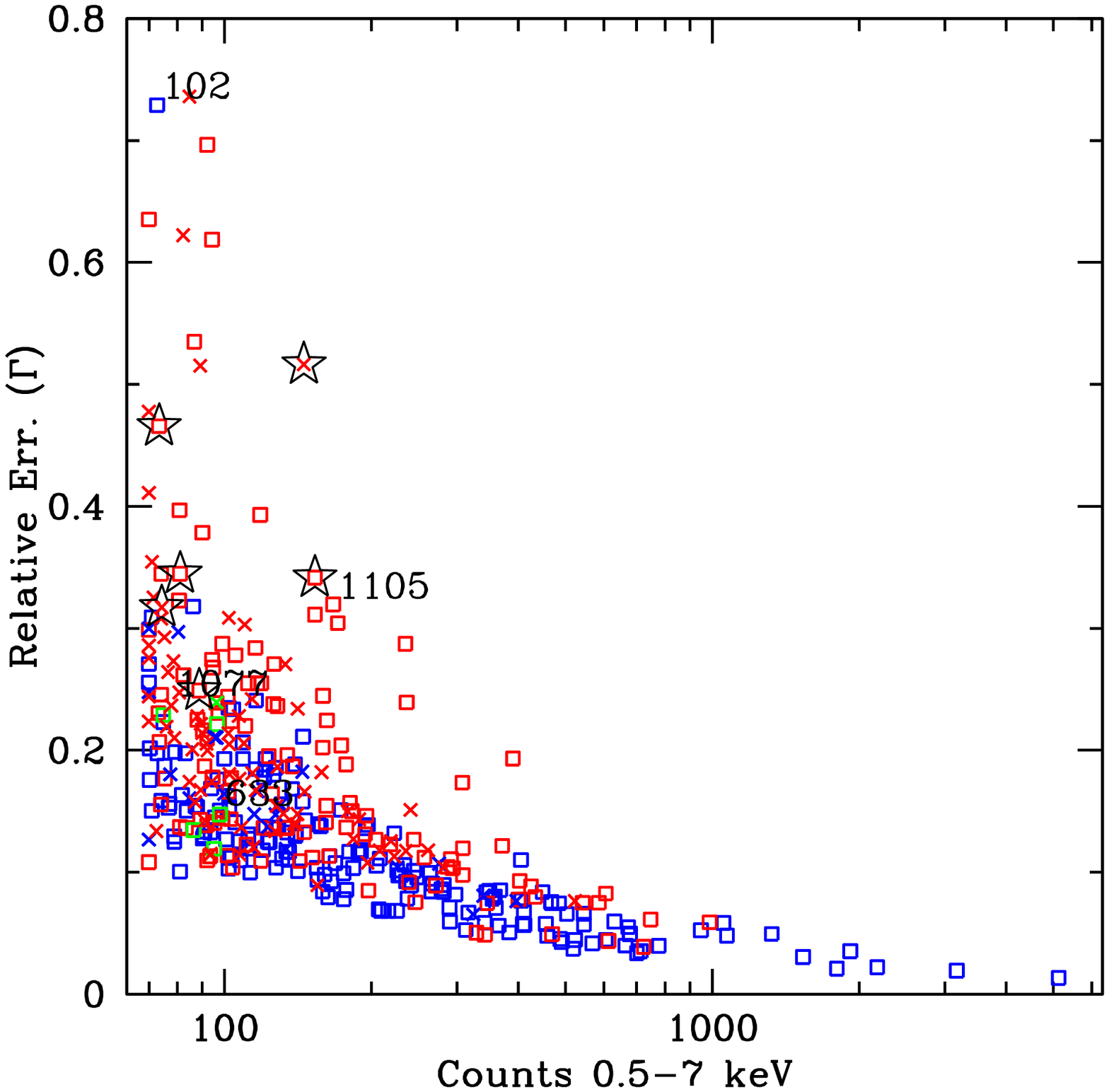}
\caption{{\it Left panel:} Distribution of $\Gamma$ as a function of the 0.5-7 keV net counts for the CCBS sources.
 Symbols as in Fig. 2.
The solid line represents the mean of the distribution, and the dotted lines show the intrinsic standard deviation (see text).
The dashed line is the  least-squares fit weighted by the errors in $\Gamma$.  
{\it Right panel:} Distribution of relative errors on $\Gamma$ with respect to 
the number of 0.5-7 keV net counts. The accuracy on the parameter determination is better than $30\%$ for $95\%$ of the sources.
In both panels peculiar sources and CT candidates are labeled as in figure 4.
}
\end{center}
\label{chan}
\end{figure*}

The global \chandra\ ACIS-I background between 0.5 and 7 keV is complex, 
and can be reproduced by two power laws, modified by photoelectric absorption, plus 3 narrow Gaussian lines 
to reproduce the features at 1.48, 1.74 and 2.16 keV (Al, Si and Au instrumental lines), 
plus a thermal (MEKAL) component.
All these components account for the sum of the particle, cosmic and galactic 
components of the background (Markevitch et al. 2003, Fiore et al. 2012).
The global background shape was recovered from background spectra extracted 
from a large (8\arcmin\ radius) regions from different pointings, 
excluding the astrophysical sources, in order to collect a large number of counts.
This allows for a detailed global characterization of the background.
The model is then rescaled (for the extraction area) to each local background and fitted, 
leaving as free parameters the overall normalization and the photoelectric absorption, 
to account for variations due to different position in the detector, which are typically more prominent in the soft (0.5-2 keV) band.
The background spectra for all the CCBS sources have very similar 
shape and normalization and the background accounts for just $\sim10\%$ percent of the total flux even at 70 net counts level.

Given the limited number of counts available, the spectral fit of the sources was performed with a simple model:
a power law, modified by photoelectric absorption at the source redshift,
plus Galactic absorption, fixed to the value in the direction of the field 
(N$_{H, Gal}$ = $2.5\times10^{20}$ cm$^{-2}$, Kalberla et al. 2005).

The distribution of the reduced Cstat (Cstat$_\nu$)
of the best fits for the 390 sources in the CCBS sample is shown in 
Fig. 3 (left panel) as a function of the number of 0.5-7 keV net counts.
The bottom panel shows the histogram of Cstat$_\nu$ for the sample.
The distribution has a peak around 1 as expected, with $\sigma= 0.20$\footnote{The distributions are different 
if the spectra are either unbinned or binned with only 1 or 5 counts per bin.
For a detailed discussion on the behavior of the Cstat minimization technique under different binning conditions,
and a comparison with $\chi^2$ statistic see Appendix A.}.

Fig. 3 (right panel) shows the distribution of Cstat with respect to DOF
for each source. The solid line represents the ideal line for a good fit (Cstat$_\nu=1$),
while the dashed lines show, for any given number of DOF, the Cstat value
above (below) which we expect a 1\% probability (one-sided) to find such 
high (low) value if the model is correct.
The quality of the fit is very good, given that almost all (383/390) the sources in our sample are distributed 
between the two dashed lines, i.e. the simple model used is in general able to satisfactorily reproduce the spectra:
We find only 7 sources ($1.7\%$ of the sample) above the upper dashed line, and 3 sources ($0.7\%$) below the lower one.
In 3 cases these sources have actually more complex spectra, requiring additional components in the model to be reproduced (see below).
The others are instead more (or less) noisy (either in the source or in the background spectra) than average.
As an example, the source with the highest value of Cstat with respect to the DOF is the source with \chandra\ ID (CID)-1105,
which has Cstat$_\nu$=1.87. This source has a complex spectrum, showing both a strong obscuration and an extra soft component.
This peculiar source, together with others are discussed in Sec. 4.3.

\begin{table*}
\begin{center}
\caption{Best fit X-ray spectral parameters for the CCBS.}
\begin{tabular}{ccccccccccccc}
\hline\hline\\
\multicolumn{1}{c} {CID} & 
\multicolumn{1}{c} {z}  &        
\multicolumn{1}{c} {Type} &          
\multicolumn{1}{c} {Flag$^a$} &
\multicolumn{1}{c} {\nh} &
\multicolumn{1}{c} {$\Gamma$} & 
\multicolumn{1}{c} {$F_{0.5-2 keV}$} &       
\multicolumn{1}{c} {$F_{2-10 keV}$} &     
\multicolumn{1}{c} {$L_{2-10 keV}$} &
\multicolumn{1}{c} {Cstat} &
\multicolumn{1}{c} {DOF} &
\multicolumn{1}{c} {Counts}\\
 &  & & & cm$^{-22}$ &  & cgs & cgs & cgs & & & \\  
\hline\\  
20  &   1.156 & Type-2 & SP & $5.77_{-2.41}^{+2.86}$ & $1.86_{-0.42}^{+0.46}$ &  1.23$\times10^{-15}$ &  6.73$\times{-15}$ & 5.49$\times10^{43}$ &  84.03 & 98 & 125.9 \\  
21  &   1.850 & Type-1 & SP &  $<0.23$ & $1.59_{-0.09}^{+0.09}$ & 7.49$\times10^{-15}$ &  1.71$\times10^{-14}$ &  4.18$\times10^{44}$ &  95.66 & 133 &  412.1 \\  
\hline                                                    
\hline
\end{tabular}
\end{center} 
$^a$ SP (PH) for spectroscopic (photometric) redshifts.
\end{table*}

\subsection{Modeling results}

Fig. 4 shows the distribution of the two main spectral parameters, $\Gamma$ and \nh, 
for the CCBS sample. There is no significant correlation between $\Gamma$ and \nh\ in the 
sample \footnote{We test for correlations using the Kendall correlation test (probability P=0.76 of no correlation) 
available in ASURV, a survival statistics package (Lavalley et al. 1992).}
even if there can be in individual sources with low statistics.
The photon index is distributed around the typical value for AGN ($\langle \Gamma \rangle =1.84$)
with observed dispersion $\sigma_{obs}=0.39$.
This dispersion is the result of the convolution of the intrinsic 
distribution with the error distribution. 
Using the maximum likelihood method of Maccacaro et al. (1988),
and assuming that both distributions are Gaussian,
we estimated an intrinsic dispersion $\sigma_{int}=0.22$, 
a very small dispersion, indicating high homogeneity in the sample spectra.
The sources with low obscuration (\nh $<10^{22}$ cm$^{-2}$) show 
a smaller dispersion in $\Gamma$ ($\sigma_{obs}=0.29$, $\sigma_{int}=0.09$), while the spread and the errors in $\Gamma$ 
increase strongly with increasing \nh.
This is due to the fact that a large amount of obscuration makes it more difficult to retrieve the intrinsic
value of the photon index.

From Fig. 4 it is possible to identify few sources with peculiar spectra,
namely three extremely soft sources, with $\Gamma>3$, (between 2 and 3$\sigma$ from the mean $\Gamma$ value), and six flat sources,
showing a photon index significantly flatter (at $3\sigma$) than the mean $\Gamma$ values,
for which a reflection dominated model may be a better description of their X-ray spectra.
All these sources are discussed in Sec. 4.3 (we also discuss there source CID-102, showing the highest 
value of \nh\ in the sample).
We excluded these sources when computing the distribution of $\Gamma$.
After excluding these outliers, the results are $\langle \Gamma \rangle =1.86$ with dispersions $\sigma_{obs}=0.34$
and $\sigma_{int}=0.14$, i.e. the intrinsic dispersion is reduced of almost 50\%.

Fig. 5 (left panel) shows the distribution of photon index $\Gamma$ as a function of the number 
of source counts for the CCBS sample (symbols as in Fig. 2).
The mean value $\langle \Gamma \rangle =1.86$ and the intrinsic standard deviation $\sigma_{int}=0.14$ 
are reported with solid and dotted lines respectively.
A least-squares fit, weighted by the errors in $\Gamma$, gives an almost flat slope (dashed line).
As expected, for lower number of counts, the spread in the values of $\Gamma$ increases, along with the errors.
The mean value of the distribution is close to the average values found for large samples 
of quasars in the 2-10 keV band (e.g. Reeves \& Turner 2000; Piconcelli et al. 2005, Tozzi et al. 2006).
Fig. 5  (right panel) shows the distribution of relative errors on $\Gamma$ with respect to 
the number of source counts.
As can be seen,  the photon index can be determined with an uncertainty smaller than
30\% in $95\%$ of the sample. 

Table 2 summarizes the results from the best fit of the X-ray spectra for each source\footnote
{The full table is available online at {\it https://hea-www.cfa.harvard.edu/$\sim$fcivano/C\_COSMOS\_X-ray\_spectral\_analysis\_CCBS.fits.}}.
We report the CID number, redshift, classification, flag for spectroscopic or photometric classification, \nh\ and $\Gamma$ with relative errors, 
0.5-2 keV flux, 2-10 keV flux, 2-10 keV luminosity, Cstat value, DOF value, number of counts.

\subsection{Peculiar sources}

In the following we discuss a small sample of peculiar sources.
Four sources have peculiar or complex X-ray spectra\footnote{Another source in the CCBS with peculiar features around the Fe $K\alpha$ line, namely CID-42, is discussed in Civano et al. (2010).}, 
while 6 sources
have very flat, possibly reflection dominated spectra.
The four sources with peculiar spectra are labeled in Fig. 4 and 5
with their CID number, while CT candidates are labeled with stars.

\begin{figure*}
\begin{center}
\includegraphics[width=4cm,height=4.5cm,angle=-90]{spec_1105.ps}\hspace{0.1cm}
\includegraphics[width=3.5cm,height=3.5cm,angle=-90]{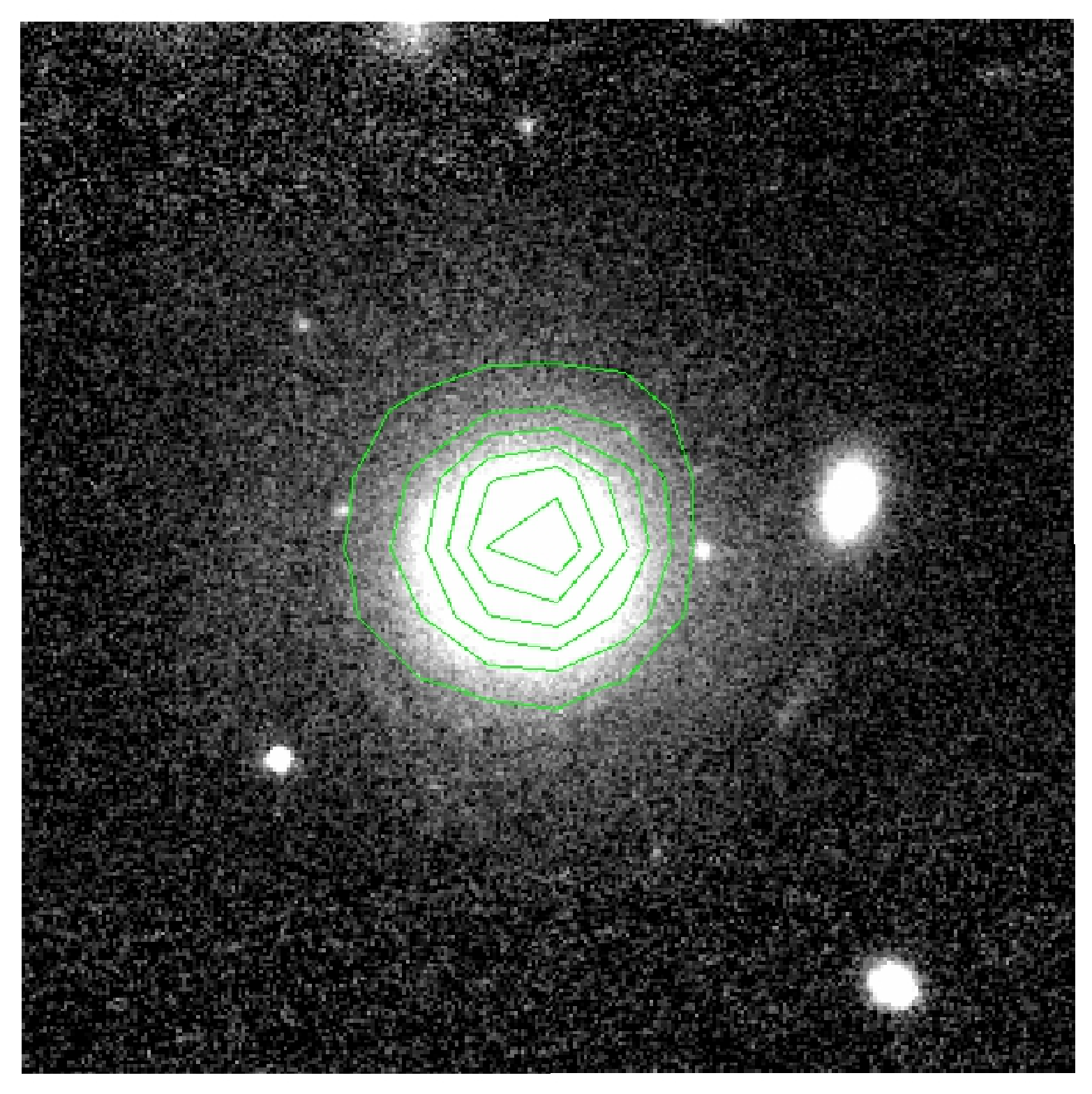}\hspace{0.5cm}
\includegraphics[width=4cm,height=4.5cm,angle=-90]{spec_633.ps}\hspace{0.1cm}
\includegraphics[width=3.5cm,height=3.5cm,angle=-90]{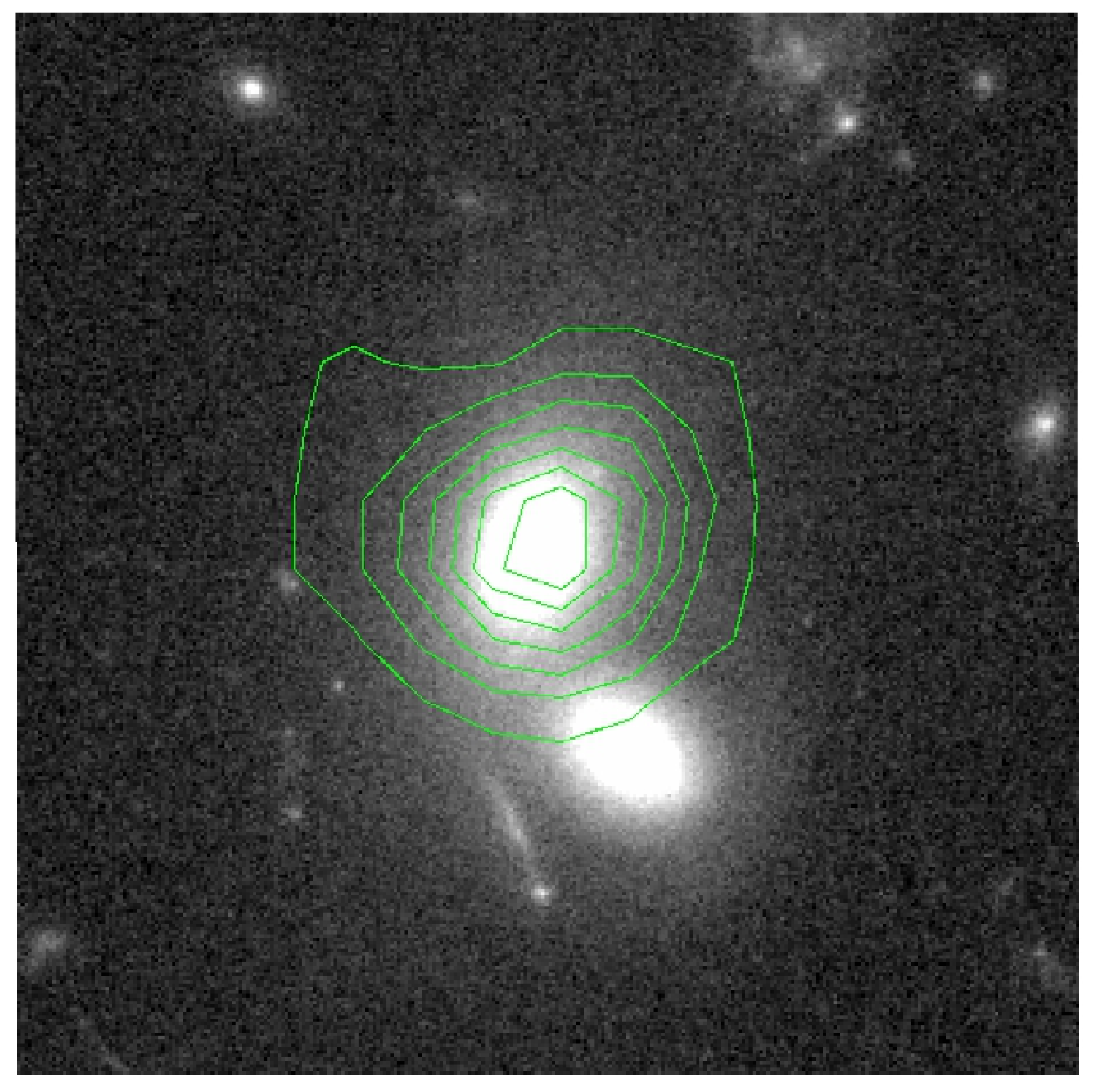}

\vspace{0.5cm}

\includegraphics[width=4cm,height=4.5cm,angle=-90]{spec_1077.ps}\hspace{0.1cm}
\includegraphics[width=3.5cm,height=3.5cm,angle=-90]{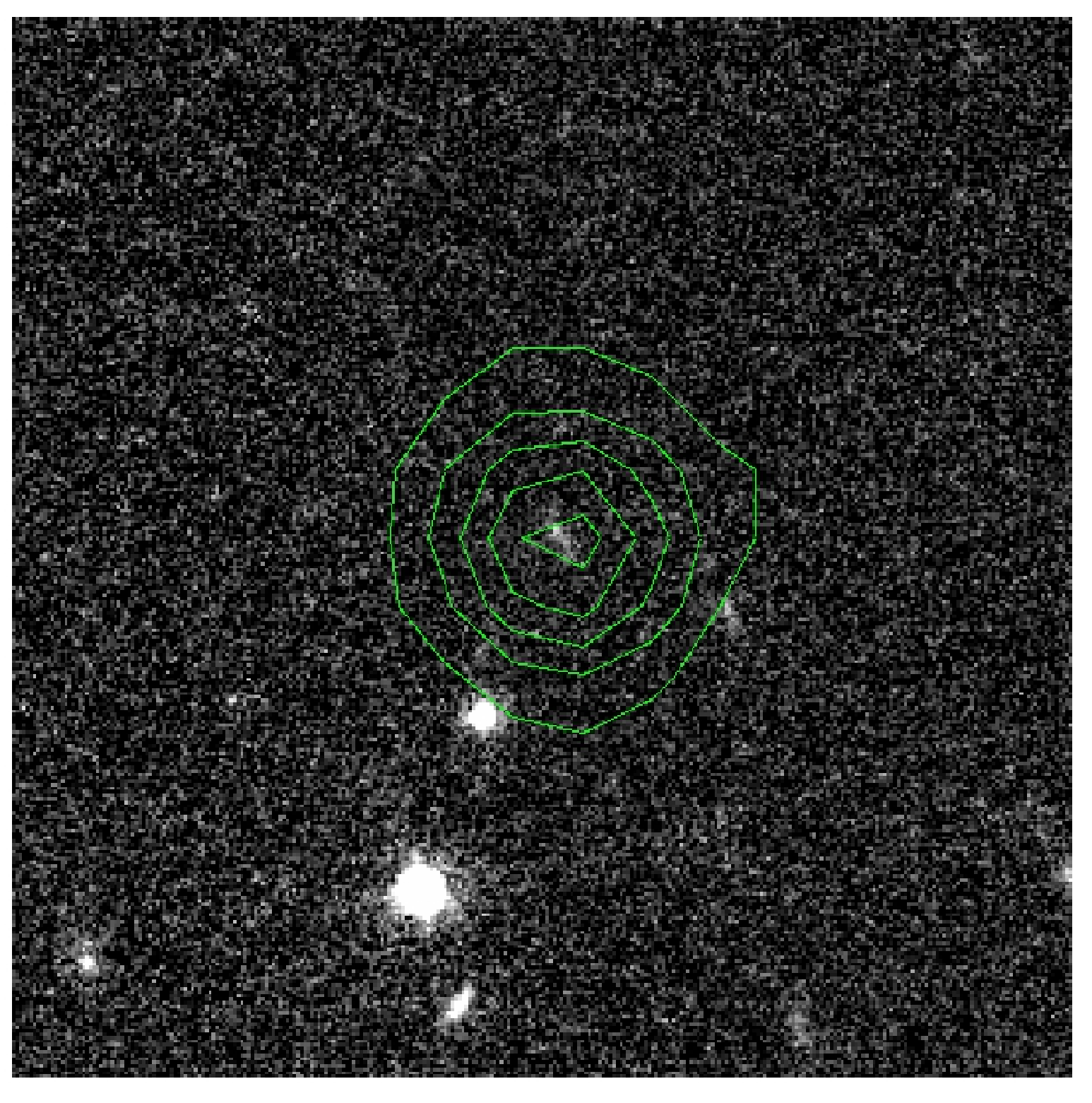}\hspace{0.5cm}
\includegraphics[width=4cm,height=4.5cm,angle=-90]{spec_102.ps}\hspace{0.1cm}
\includegraphics[width=3.5cm,height=3.5cm,angle=-90]{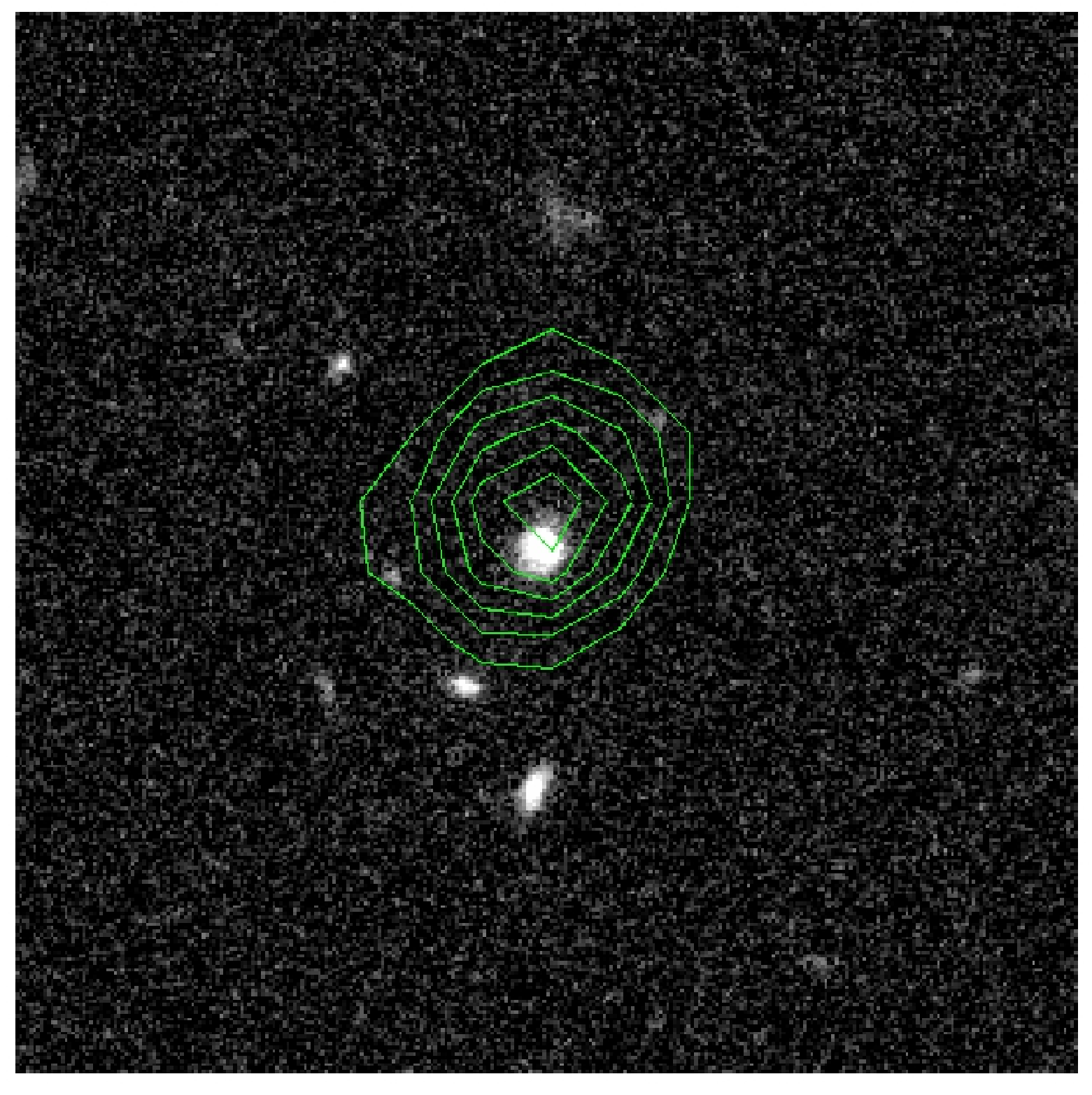}
\caption{\chandra\ 0.5-7 keV  spectra (left panel) and {\it Hubble}-ACS image (right panel) of four peculiar sources in the CCBS: 
CID-1105, 633, 1077 and 120, from top left to bottom right.
The total spectrum is shown in black, and the background spectrum in red, each with its best fit model. 
The counts are binned to a minimum significant detection of $3\sigma$ 
(and a maximum of 15 counts per bin) for plotting purposes only. The lower part of each left panel shows the model to data ratio for the total spectra only.
In the right panels the \chandra\ ACIS-I contours (in green) are superimposed to the ACS images.}
\end{center}
\label{peculiar}
\end{figure*}

\begin{figure*}
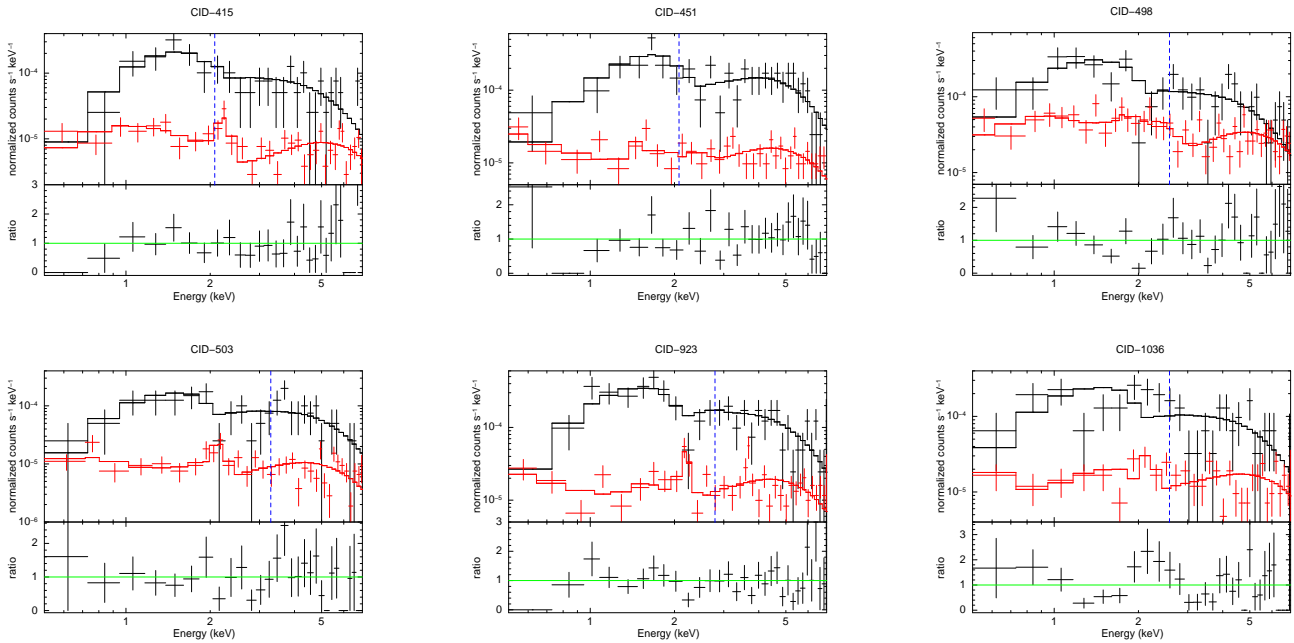

\begin{center}
\includegraphics[width=4cm,height=5cm,angle=-90]{spec_415.ps}\hspace{1.1cm}
\includegraphics[width=4cm,height=5cm,angle=-90]{spec_451.ps}\hspace{1.1cm}
\includegraphics[width=4cm,height=5cm,angle=-90]{spec_498.ps}\hspace{1.1cm}

\vspace{0.1cm}

\includegraphics[width=4cm,height=5cm,angle=-90]{spec_503.ps}\hspace{1.1cm}
\includegraphics[width=4cm,height=5cm,angle=-90]{spec_923.ps}\hspace{1.1cm}
\includegraphics[width=4cm,height=5cm,angle=-90]{spec_1036.ps}\hspace{1.1cm}
\caption{\chandra\ 0.5-7 keV  spectra of the CT candidates in the CCBS: CID-415, 451, 498, 503, 923 and 1036 from left to right and top to bottom.
The total spectrum is shown in black, and the background spectrum in red, each with its best fit model. 
The counts are binned to a minimum significant detection of $3\sigma$ 
(and a maximum of 15 counts per bin) for plotting purposes only.
The lower part of each panel shows the model to data ratio for the total spectra.
The blue dashed line represents for each source the location of the expected 
Fe K$\alpha$ emission line at 6.4 keV.}
\end{center}
\label{ctcandidates}
\end{figure*}

\subsubsection{CID 1105}

This source at redshift z=0.219 shows the highest value of Cstat with respect to DOF 
(Cstat$_\nu$=1.87).
The source also shows the second highest value of \nh\ in the sample,
and a steep power-law.
As can be seen in Fig. 6 (top left panel), the source has a complex spectrum, showing a strongly obscured primary power-law, and an extra soft component
superimposed to it.
The soft component can be modeled either with a thermal component (kT$=0.20_{-0.05}^{+0.09}$ keV at the source redshift),
or with a very soft, second power-law ($\Gamma=3.8_{-1.2}^{+1.4}$).
Both fits with an additional components are significant improvements with respect to the simple absorbed power law fit,
having a Cstat$_\nu$=0.99 and 0.98, respectively. The F-test 
performed to check if the addition of an extra component
is required, gives probabilities of $P=4.23\times10^{-13}$ and $P=1.53\times10^{-11}$  respectively, i.e. the extra component is required 
by the data at very high significance. Given the quality of the data available, it is not possible to establish which one of the two models is preferred by the fit,
even if the use of a thermal component results in a slightly better fit.
Furthermore, the optical {\it Hubble}-ACS image shows a bright, extended elliptical galaxy.
This suggests that the source is an obscured AGN, with the soft X-ray component coming from thermal emission from the intra-galactic gas,
with possible contribution from the AGN scattered component.
The results of both fits with additional soft components are consistent 
($\Gamma=2.42_{-0.47}^{+0.76}$, \nh $=44.0_{-15.1}^{+20.3}$ cm$^{-2}$, $L_{2-10 keV}=3.7\times10^{43}$ erg s$^{-1}$) 
and will be adopted in the rest of the discussion and in Table 2.

\subsubsection{CID-633}

This is the only one source in the CCBS which shows a steep spectrum ($\Gamma=3.1_{-0.41}^{+0.49}$) and a low value of \nh.
The \chandra\ spectrum of the source, which is optically classified as a non active galaxy at z=0.35,
is shown in Fig. 6 (top right panel) and 
can be reproduced with a thermal emission with temperature $kT = 0.35_{-0.05}^{+0.06}$ keV at the source redshift,
compatible with emission from hot gas in the galaxy. 
This is also consistent with 
a relatively low X-ray luminosity of L$_{2-10 keV}\sim10^{42}$ erg s$^{-1}$.
The ACS image shows that this source has a neighbor, with the two nuclei being at 3.15\arcsec.
The X-ray emission corresponds to the central region of CID-633.
We conclude that it is compatible with emission 
from a recent star formation event in the central region of CID-633, likely triggered by the interaction with the neighbor galaxy.
This source is excluded from the following discussion on the power-law fit parameter distribution.

\subsubsection{CID-1077}

This source at z=2.017 has as best fit SED a galaxy template 
from Ilbert et al. 2009.
It has a 2-10 keV luminosity of $3\times10^{43}$ erg cm$^{-2}$ s$^{-1}$.
The \chandra\ spectrum of the source (Fig. 6, bottom left panel)
shows a high column density (of the order of $10^{23}$ cm$^{-2}$) and a steep spectrum
($\Gamma=3.42_{-0.72}^{+0.90}$). 
This combination of spectral parameters is quite unusual, and can be due to a thermal spectrum,
like in the case of CID-633, if the photometric redshift for this source were incorrect (few \% of outliers are expected).
However, from the ACS image, it appears that the optical counterpart is very faint, compatible with
the high redshift scenario. In this case it is hard to explain the steep power-law characterizing the \chandra\ 
spectrum of this source.

\subsubsection{CID-102}

This source has the highest value of \nh\ in the CCBS sample (\nh$=4.4_{-2.4}^{+3.6}\times10^{23}$ cm$^{-2}$)
and a flat spectrum ($\Gamma=1.30_{-0.85}^{+1.04}$), but is optically classified as a broad line AGN
at z=1.84. 
However, the X-ray emission of this source is just above our selection limit.
The X-ray spectrum (Fig. 6 bottom right panel) has 70.2 net counts (0.5-7 keV) and shows a hint of
a second soft component below 1 keV. As a result, the source shows one of the highest values of reduced Cstat
(Cstat$_\nu$=1.36) and high values of relative errors 
(the source is the second from the top in Fig. 5 right panel).
Adding a soft component (either a second power-law or a black body component)
to the model produces a significant improvement in the fit (Cstat$_\nu\sim1.15$ in both cases),
giving similar results on \nh\ and $\Gamma$ of the previous model, but with smaller errors.
We will use these results in the following discussion.
However, given the quality of the spectrum, we cannot draw any firm conclusions for this peculiar source.

\subsubsection{Compton Thick candidates}

\begin{figure*}
\begin{center}
\includegraphics[width=8cm,height=8cm]{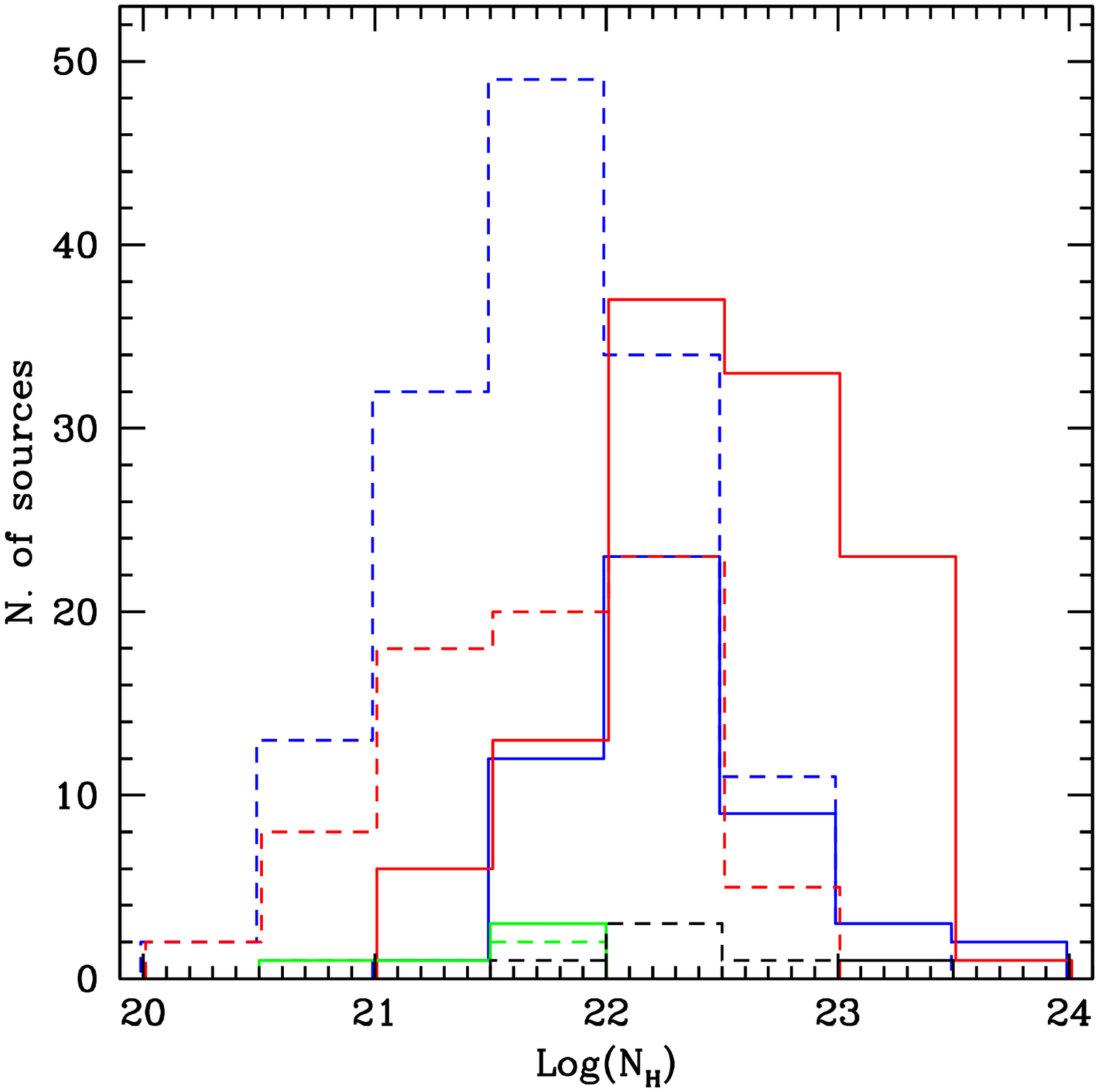}\hspace{1.5cm}\includegraphics[width=8cm,height=8cm]{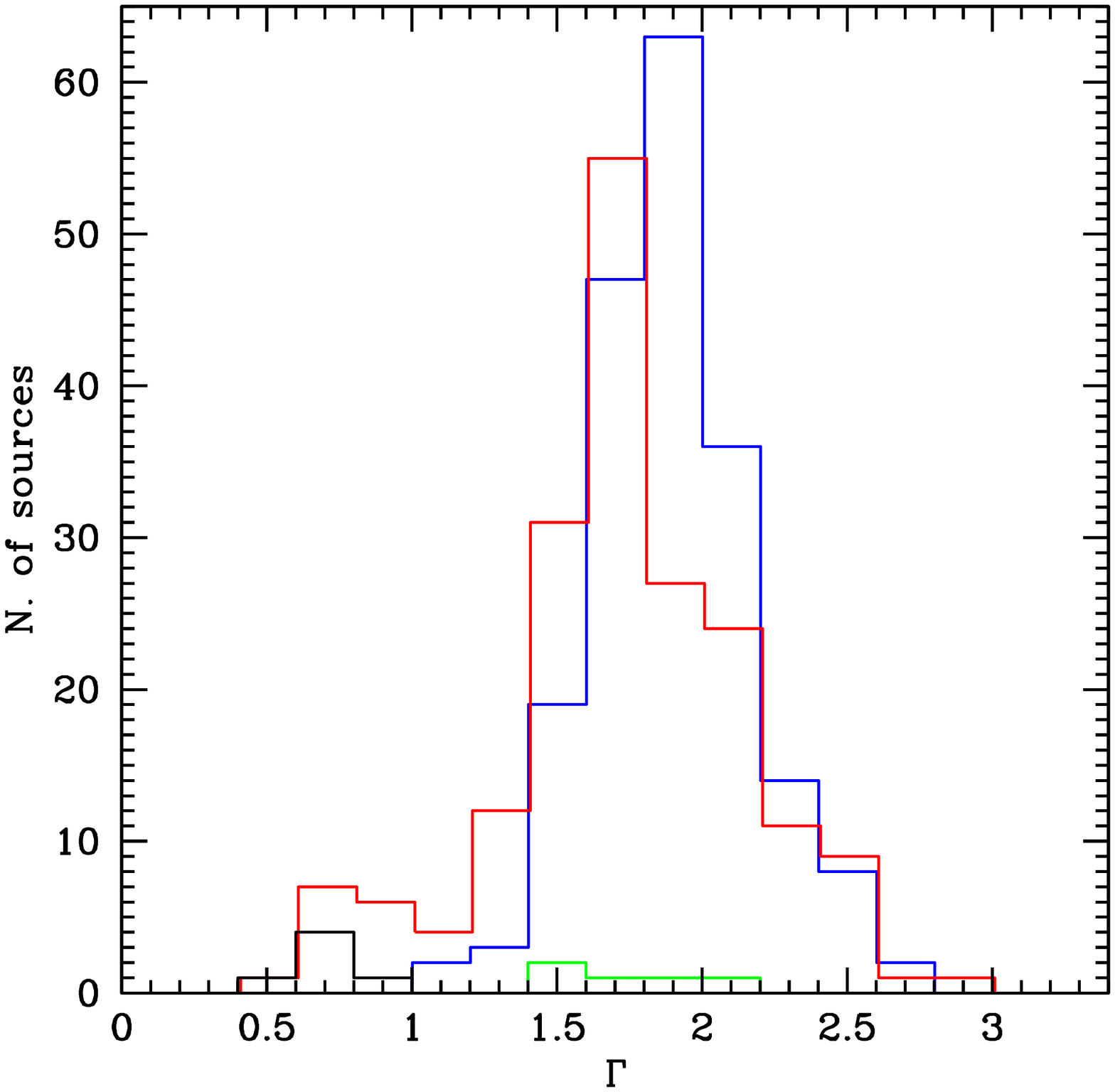}
\caption{{\it Left panel:} Distribution of column density \nh~. In blue are represented type-1 AGN, in red type-2 AGN, 
in green sources classified as galaxies, in black CT candidates. The solid (shaded) histograms represent sources with a detection (an upper limit) of \nh.
{\it Right panel:} Distribution of photon index $\Gamma$. Colors as in left panel.}
\end{center}
\label{histonh}
\end{figure*}

There is a small population of sources with very flat 
spectra, for which a reflection dominated, CT model 
may be a better description of the extremely flat spectrum.
The ``smoking gun" signature in the spectra of these highly obscured AGN would be a strong (EW$\sim1$ keV) 
Fe K$\alpha$ fluorescent emission line at 6.4 keV (George \& Fabian 1991; Matt, Perola, \& Piro 1991).
However, almost all of these sources have less than 100-200 counts (Fig. 5 left), 
making difficult to detect even such a strong line.

The spectra of sources which show a photon index significantly flatter than the average $\Gamma$,
namely CID-415, 451, 498, 503, 923 and 1036, are shown in Fig. 7 from top left to bottom right.
The blue dashed line represent for each source the location of the expected 
Fe K$\alpha$ emission line at 6.4 keV.
A very flat power law, in some case with the addiction of significant (but always $<3\times10^{23}$ cm$^{-2}$) obscuration, can reproduce 
the shape of the spectra of these hard sources. However, a completely reflected power law (i.e. {\it xspexrav} model in Sherpa, based on the 
{\it pexrav} model of XSPEC, with pure reflection) can give similar results but, given the low data quality, 
it is not possible to determine the preferred model on the basis of the fit results.
A strong ($EW\sim1$ keV) Fe K$\alpha$ line at 6.4 keV is expected as part of the reflected emission,
but no sign of such signature is present in four of the observed spectra, with the possible exception of sources CID-503 and CID-1036.
The upper limit on the equivalent width (EW) of the emission line is in the range 0.09-0.38 keV, depending on the quality of the data,
with average value $\langle EW \rangle=0.20$ keV.

The source CID-503 shows some residuals at energies just above 6.4 keV, possibly indicating the presence of an Fe K$\alpha$ emission line.
If the energy and normalization of the line are left free to vary (with fixed $\sigma=0.01$ keV) we obtain an emission line at energy 
E$_{line}=6.93_{-0.11}^{+0.11}$ keV. 
The equivalent width is EW$=0.65_{-0.21}^{+0.25}$ keV. 
The energy of the line is compatible, within $2\sigma$, with a ionized Fe line at 6.7 keV,
that appears to increase in importance in high redshift type-2 AGN (Iwasawa et al. 2012).

For source CID-1036 the best fit is obtained if also the width of the line is left free to vary:
a broad ($\sigma=0.3$ keV) emission line at energy E$_{line}=5.85_{-0.52}^{+0.58}$ keV, but compatible with 6.4 keV within one 
$\sigma$, is required. We underline that the photometric redshift of the source  ($z=1.497$) has an error of $\pm0.02$ at 1$\sigma$.
The EW is compatible to the one expected from a reflection dominated spectrum (EW$=1.20_{-0.65}^{0.93}$ keV).

We also computed the expected number of CT in our sample, according to the most recent 
predictions of X-ray background synthesis models (Gilli et al. 2007, Treister et al. 2009).
Using the {\it pexrav} model of pure reflection we translated the 70 counts (0.5-7 keV) cut into a flux limit of 
$\sim1.1\times10^{-14}$ erg cm$^{-2}$ s$^{-1}$
in the 2-10 keV band (for a nominal exposure time of 160 ks).
The Gilli et al. model\footnote{http://www.bo.astro.it/~gilli/counts.html} 
predict $\sim4$ CT sources per deg$^{-2}$ above this flux limit. 
This, folded with the C-COSMOS area-sensitivity curve, translates into $2-3$ CT sources expected.
Therefore the lack of confirmed CT sources in our sample is not in disagreement with predictions from these
background synthesis models, and is consequence of the flux limit of our sample.

\begin{figure*}
\begin{center}
\includegraphics[width=8cm,height=8cm]{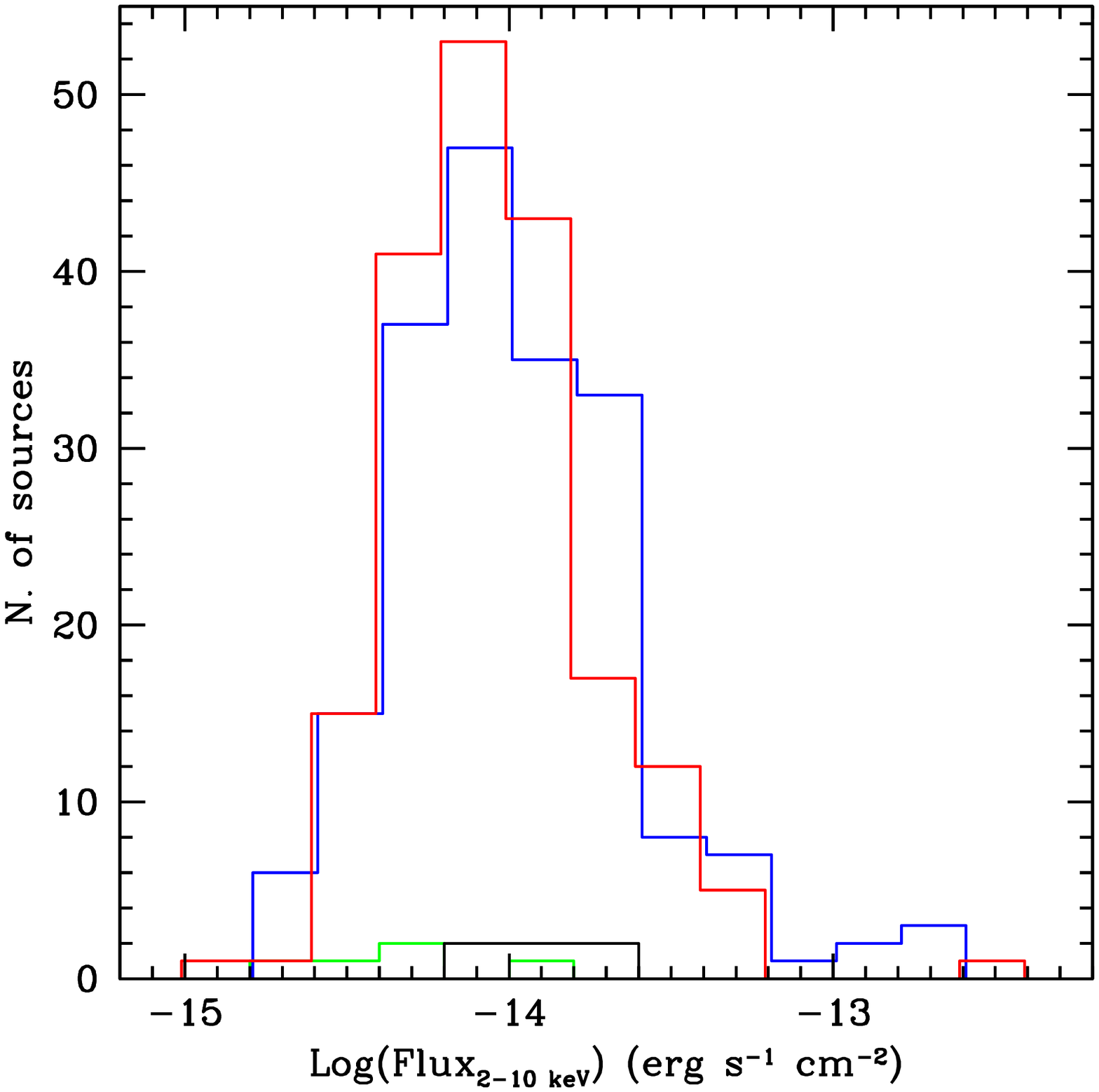}\hspace{1.5cm}\includegraphics[width=8cm,height=8cm]{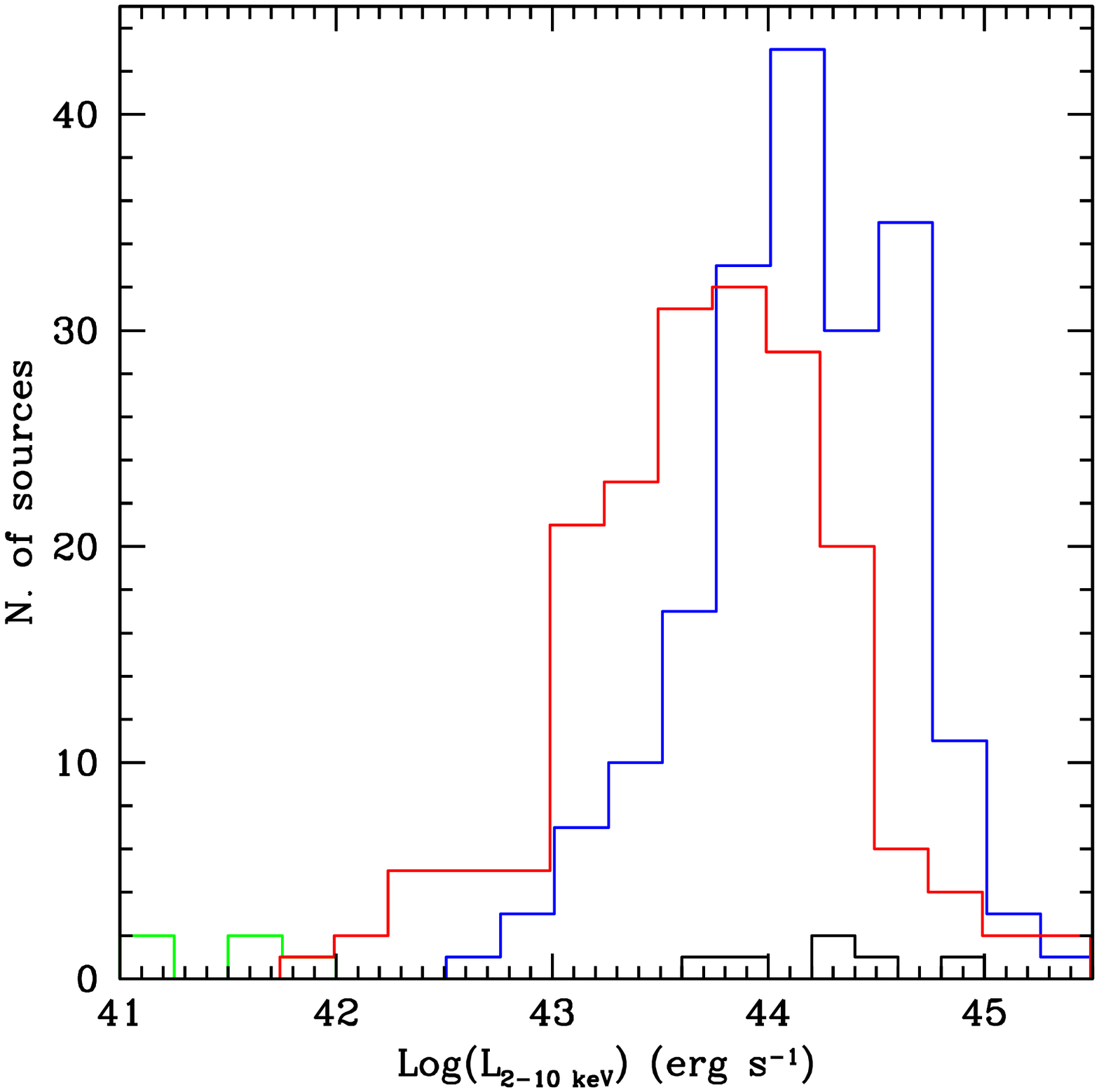}
\caption{Distribution of 2-10 keV observed flux (left) and 2-10 keV intrinsic luminosity, corrected for absorption (right). 
Colors as in previous figures.}
\end{center}
\label{histolum}
\end{figure*}

\section{Fit parameter distributions}

Having verified that our results are statistically robust (see also Appendix A), we can 
discuss the distribution of the derived spectral parameters.
In Fig. 8 we show the distributions of two spectral parameters \nh~and $\Gamma$ 
(left and right panels, respectively). 
In both plots colors identify different classes of sources, blue for type-1, red for type-2, green for galaxies and
black for the sample of CT candidates.
In the left panel, the solid (shaded) histograms represent detections (upper limits) of \nh. 

\subsection{\nh, intrinsic absorption column density}

For type-1 AGNs, the distribution of intrinsic obscuration has a global peak in the bin between 
$3-10\times10^{21}$ cm$^{-2}$, but is strongly dominated by upper limits (75\% of the sources).
The fraction of sources that can be considered obscured (\nh$>10^{22}$ cm$^{-2}$), assuming a Gaussian distribution 
for the errors, and computing for each source the fraction of the Gaussian above $10^{22}$ cm$^{-2}$,
is 17\% (33 sources).
This fraction is slightly larger of what found in shallower surveys (Brusa et al. 2003; Silverman et al. 2005; 
Mateos et al. 2010; Corral et al. 2011; Scott et al. 2011),
and similar to what found for deeper surveys, like the 1 Msec. observation in the CDFS (Tozzi et al. 2006).

For type-2 AGN instead, the situation is inverted. The distribution is peaked above $10^{22}$ cm$^{-2}$, and
60\% of the sources have a detection of intrinsic column density.
However, the fraction of type-2 that can be considered obscured (computed as for type-1s),
is only 50\%. 
This is a low value for this class of sources, compared with previous results.
This may be due to the fact that our sample is biased against low counts sources, 
and hence in favor of less obscured ones (e.g. at F$_{2-10 keV}\sim5\times10^{-15}$ a source with 
\nh$=5\times10^{22}$ cm$^{-2}$ is excluded by our counts based selection, while an unobscured source is not).
Furthermore $\sim35\%$ of non obscured type-2s are classified on the basis of their SED fitting,
typically as passive galaxies, and not from a proper optical spectrum.
On the other hand, $\sim40\%$ of these unobscured type-2s have L$_{2-10 keV}<2\times10^{43}$ erg s$^{-1}$, 
i.e. some of them can have a significant contribution from a starbursts in the soft band.
Finally, some unobscured type-2s may be faint type-1 AGNs, 
with strong optical/IR contamination from host galaxy light, so that the SED fitting fails to reveal 
their intrinsic nature.

All the sources classified as galaxies have low upper limits in column density, consistent with Galactic values.
This is compatible with the fact that the X-ray emission of these sources comes from the integrated emission
of X-ray binaries, plus diffuse hot gas (Fabbiano 1989). The emitting region therefore is extended, not nuclear, and 
no global obscuration is expected, unless the galaxy is observed perfectly edge-on and the whole disk is obscured by a dust lane.

\subsection{$\Gamma$, photon index}

In the right panel of Fig. 8 the distribution of $\Gamma$ for the CCBS is shown. 
The distribution of $\Gamma$ for type-1 AGN is strongly peaked in the bin between 1.8 and 2, with a mean value 
$\langle \Gamma \rangle= 1.89\pm 0.02$ and $\sigma_{obs}=0.28$ . 
For type-2 AGN the distributions has a peak at slightly lower values, $\Gamma= 1.6-1.8$, 
with a mean value $\langle \Gamma \rangle= 1.76 \pm 0.02$, and $\sigma_{obs}=0.38$.
The observed dispersion for type-2 is larger than for type-1, 
however type-2 have typically larger errors on $\Gamma$, due to higher values of \nh.
Indeed, the intrinsic dispersion that we estimate taking into account the error distribution are
very similar to each other ($\sigma_{int}=$ 0.11 and 0.13 for type-1 and type-2 respectively).

A 2-sided Kolmogorov-Smirnov (KS) test gives a probability $P=4.04\times10^{-5}$ that the two distributions are drawn from the same population.
This difference can be due to a residual obscuration unaccounted in the fit,
or to the contribution of a reflection component, that is known
to increase in importance with decreasing luminosities (Iwasawa \& Taniguchi 1993);
indeed type-2 have on average lower luminosities than type-1 in the CCBS (see Fig. 9 right panel).
The average quality of our spectra does not allow to test this hypothesis
using a model that includes Compton reflection. 
Bianchi et al. (2009) verified, on the much higher quality spectra of the CAIXA AGN sample
(mostly local) that the inclusion of a luminosity-dependent reflection component does not
affect the differences between the two populations.

Among the galaxies, only one shows a very soft spectrum, compatible with thermal emission from hot gas,
while the others have spectra that are indistinguishable from those of AGN for what concern the photon index.
This can be the result of the super imposition of different components, as mentioned above.

\begin{figure*}
\begin{center}
\includegraphics[width=8cm,height=8cm]{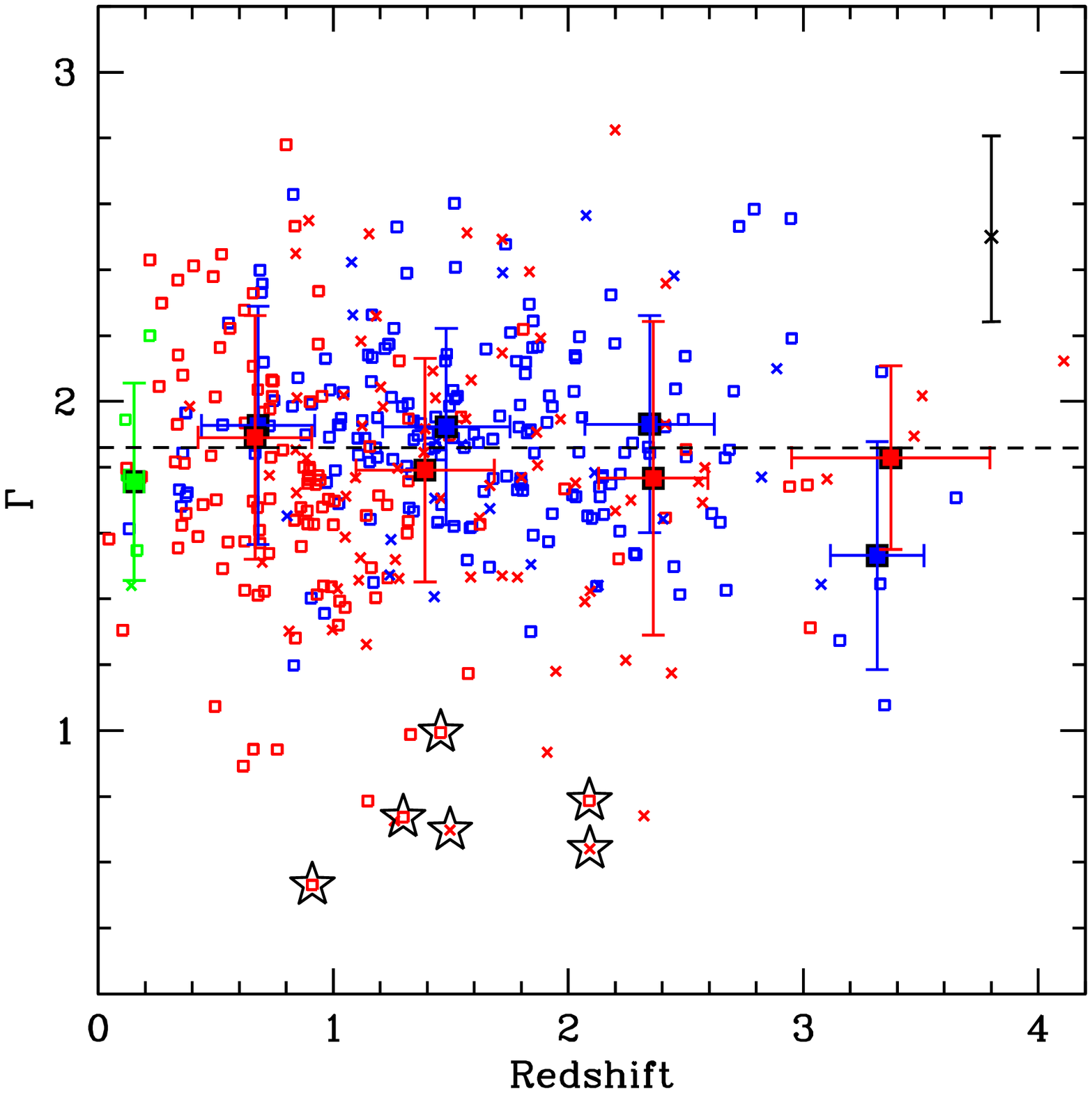}\hspace{1.5cm}
\includegraphics[width=8cm,height=8cm]{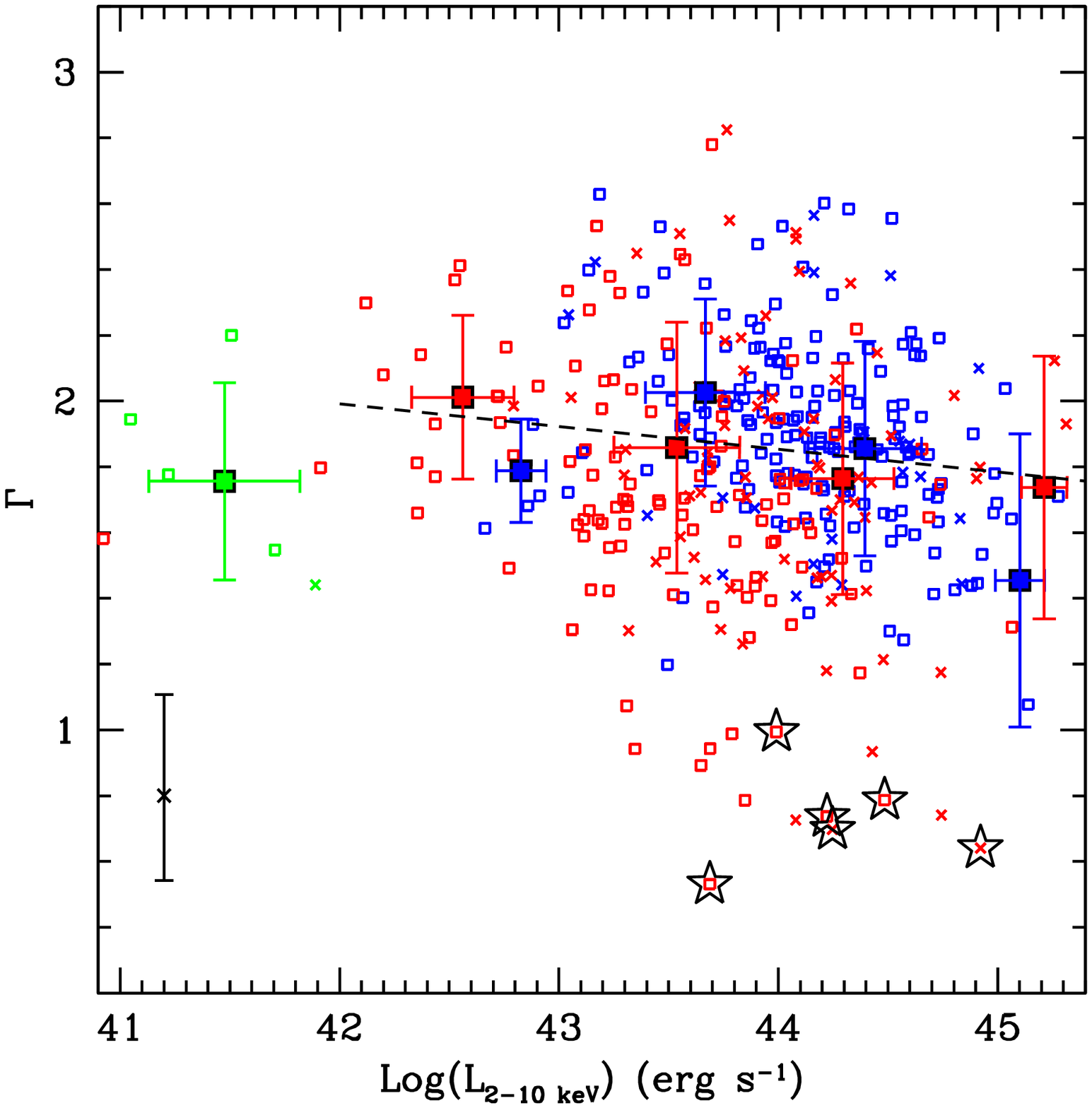}
\caption{{\it Left panel:} Distribution of photon index as a function of redshift for the CCBS sources. 
{\it Right panel:} Distribution of $\Gamma$ as a function of the 2-10 keV luminosity, corrected for absorption, for our sources.
In both panels, small symbols represent different classes of sources as in previous figures. 
The large thick squares represent the weighted mean for bins of redshift ($\Delta z$=1) and $\Delta Log(L_{2-10keV})=1$, respectively.
The error bars mark the 1$\sigma$ dispersions in both axes.
The dashed lines represent the least-squares fit to the type-1 and type-2 AGN together (CT candidates are excluded from the fit).}
\end{center}
\label{gammaz}
\end{figure*}

\subsection{Flux and Luminosity}
 
Fig. 9 (left panel) shows the distribution of the 2-10 keV band observed flux
for the CCBS sources.
The flux distributions for type-1 and type-2 are very similar  (KS probability $p=0.13$) and broad, as 
expected from the flux limited nature of our sample, with similar average values 
$\langle F_{2-10 keV} \rangle \sim 1.6\times10^{-14}$ and standard deviation $\sigma \sim 2.4\times10^{-14}$ erg cm$^{-2}$ s$^{-1}$.

Fig. 9 (right panel) shows the distribution of intrinsic 2-10 keV luminosity, corrected for absorption.
We computed the intrinsic 2-10 keV luminosity from the best fit model and the redshift of each source, 
imposing the intrinsic and Galactic absorption to be zero.
Type-1 AGN are typically more luminous in the 2-10 keV band, with 63\% of them being in the quasar regime, 
i.e. L$_{2-10 keV} > 10^{44}$ erg s$^{-1}$.
Type-2 AGN are less luminous, with only 33\% of the sources having L$_{2-10 keV} > 10^{44}$ erg s$^{-1}$.
The two distributions are significantly different (KS test probability $p=1.58\times10^{-10}$).
This difference
is the result of the different redshift distributions (type-1 AGN typically have higher redshift)
coupled with the fact that our sample is flux-limited, 
i.e. redshift and luminosities are tightly correlated (see Fig. 2).

\subsection{Photon Index dependencies}

We studied the possible evolution of the photon index with redshift and X-ray luminosity.
Fig. 10 (left panel) shows the distribution of $\Gamma$ a function of redshift, while the right panel
shows the distribution of $\Gamma$ as a function of the 2-10 keV intrinsic luminosity, corrected for absorption. 
Individual sources are color coded as in Fig. 4. The thick filled squares represent 
the weighted mean for bins of redshift ($\Delta$z=1) and luminosity ($\Delta$Log(L$_{2-10 keV}$)=1) respectively. 
The error bars mark the 1$\sigma$ dispersions in both axes in the two plots.
The CT candidates sources are marked with stars, and have been excluded from the least-square fit and
mean.

The dashed line in both panels represents the least-squares fit, weighted by the errors in $\Gamma$,
for type-1 and type-2 AGN together. We excluded from the fit sources classified as galaxies.
There is no evident correlation in redshift,
apart for a hint of harder spectra for type 1 AGN at $z>3$ (the average is at $\sim1\sigma$ from the typical value).
However we stress that there are only 6 type 1 AGN at  $z>3$, therefore we do not have enough statistic to prove that 
the effect is real.
This could be due to the fact that at these high redshifts, the reflection hump enters the observing band 
(30 keV at $z=3.5$ correspond to 6.7 keV) flattening the observed spectrum.
At the average luminosity of the sources in this bin ($\langle L_{2-10keV}\rangle=6.8\times10^{44}$ erg s$^{-1}$)
the EW of the Fe K$\alpha$ is in the range $EW=30-60$ eV (Bianchi et al. 2007) and
the contribution of the reflected component is only $\sim10\%$ in the 2-7 keV observed band.
We simulated spectra with the appropriate reflection component superimposed to the primary power-law (with $\Gamma=1.9$) as underlying model,
with $\sim1000$ and $\sim100$ counts (1000 spectra for each counts regime), 
and applied the same spectral analysis described in Sec. 4.1.
The average photon index resulting from a fit with a simple power-law hare $\langle \Gamma \rangle=1.58\pm0.08$ 
and $1.62_{-0.25}^{+0.27}$, respectively. Thus the observed hardening of the average $\Gamma$ at $z>3$ is consistent with
being produced by an unaccounted reflection component.
However, if this is the correct explanation for that effect, 
it doesn't' explain why this effect should be present for type 1s and not for type 2s.
We also underline that we only have 6 type 1s and 7 type 2s at such high redshift, therefore we
do not have enough statistics to draw any firm conclusion.

\begin{figure*}
\begin{center}
\includegraphics[width=8cm,height=8cm]{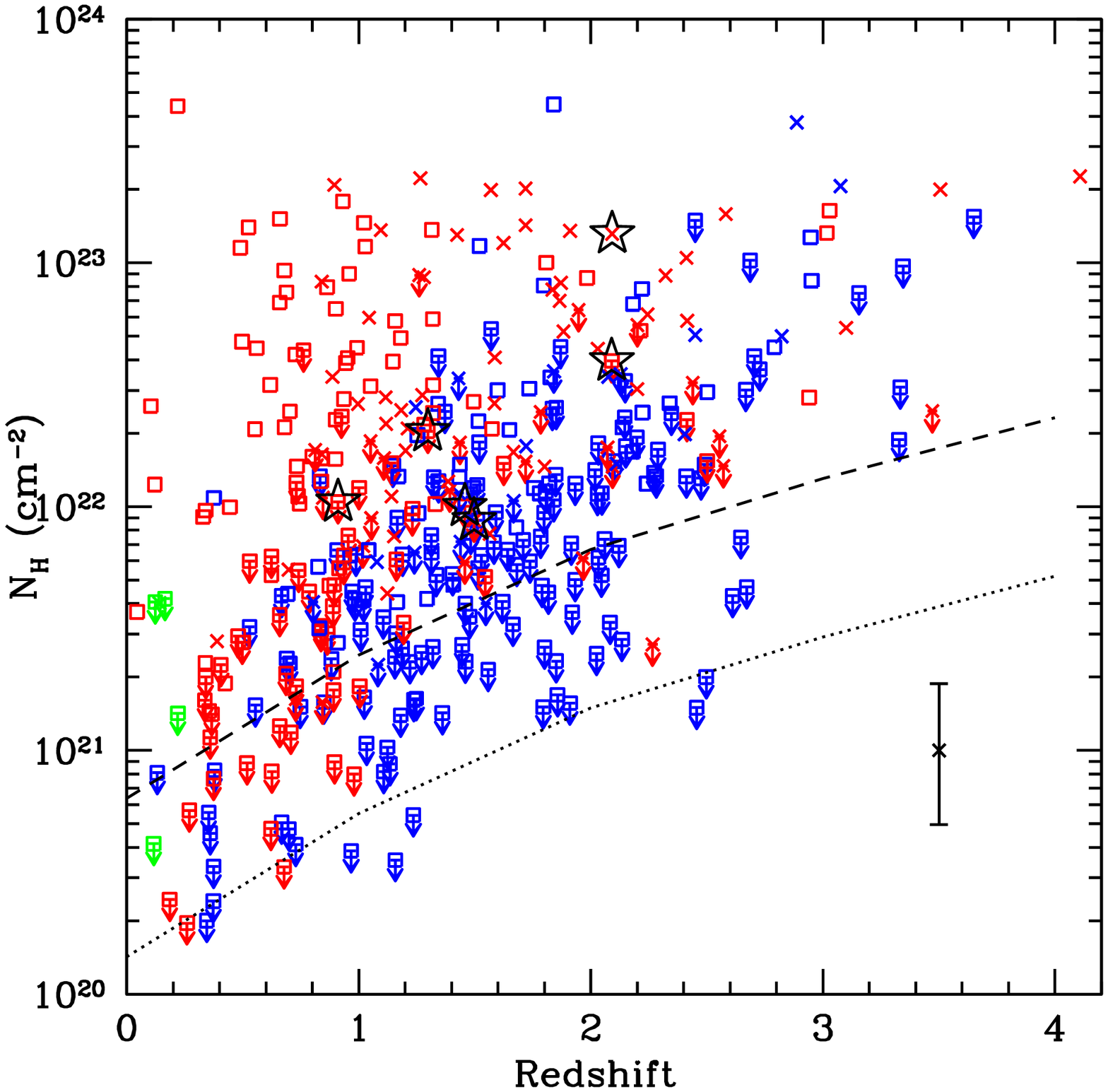}\hspace{1.5cm}
\includegraphics[width=8cm,height=8cm]{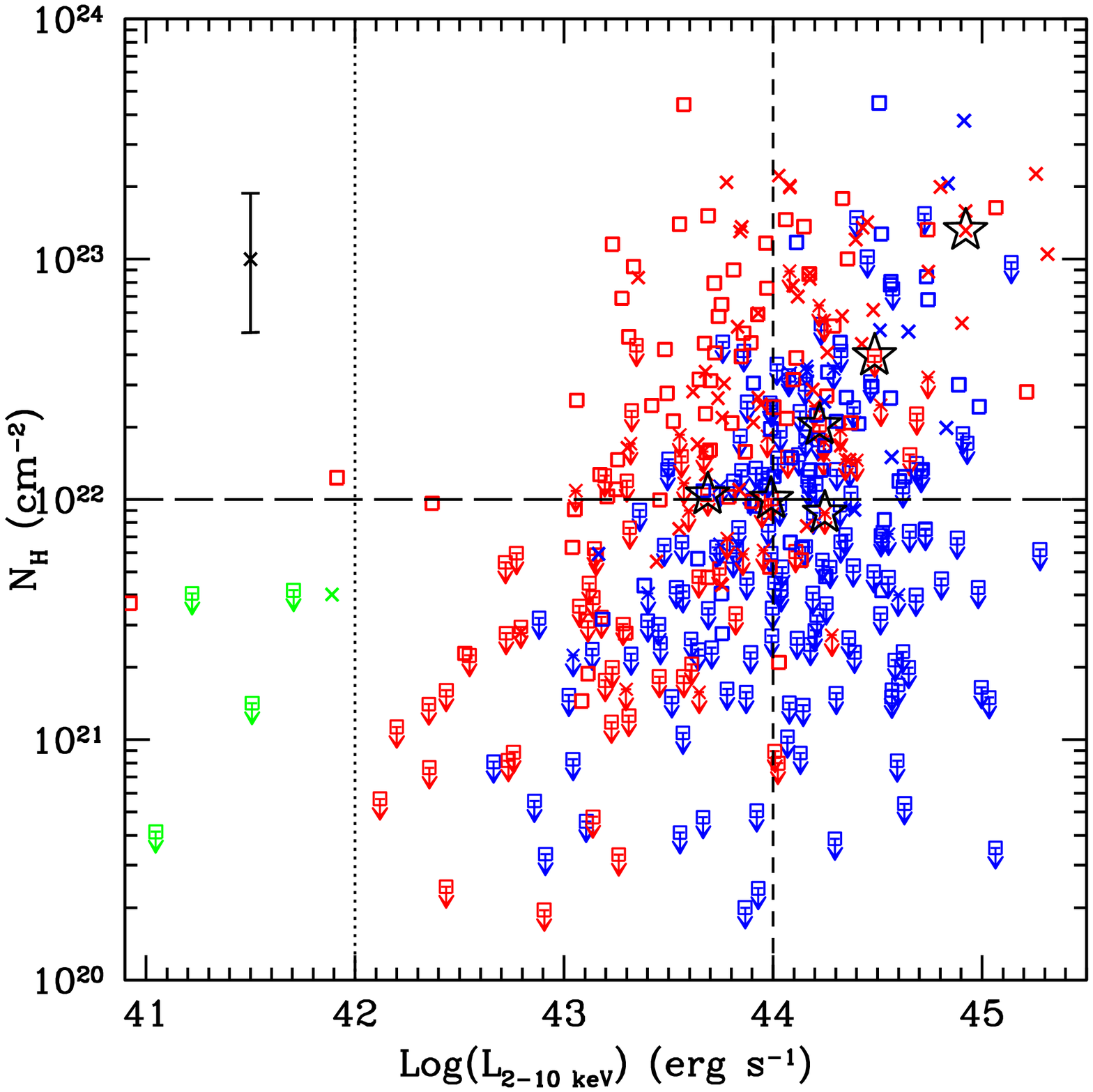}
\caption{{\it Left panel:} Distribution of intrinsic absorption, as a function of the source redshift for the CCBS. 
Symbols as in previous figures. 
The dashed (dotted) curve represents the upper limit obtained from the spectral fit of a 200 (1000) counts simulated spectrum with no intrinsic absorption
and photon index $\Gamma=1.9$, for a given redshift. 
{\it Right panel:} Distribution of intrinsic absorption, as a function of the 2-10 keV intrinsic luminosity, corrected for absorption.  Symbols as in previous figures.
The dashed line marks the separation between quasars and Seyfert (L$_{2-10 keV}=10^{44}$ erg s$^{-1}$),
the long dashed line represents the \nh\ value used to separate obscured and unobscured sources (\nh$=10^{22}$ cm$^{-2}$)
while the dotted line shows the luminosity cut we adopted to distinguish between AGN and star forming galaxies (L$_{2-10 keV}=10^{42}$ erg s$^{-1}$).
}
\end{center}
\label{nhz}
\end{figure*}

In the $\Gamma$ vs L$_{2-10 keV}$ plot a weak anti-correlation can be seen in both type-1 and type-2 classes, 
with a slope of $-0.09\pm0.06$ (but consistent with no correlation within $1.5\sigma$).
A steeper anti--correlation between photon index and hard X-ray luminosity has been found in previous work 
on \chandra\ and \xmm\ counterparts of SDSS quasars (Young et al. 2009, Green et al. 2009, Kelly et al. 2009).
On the other hand, no correlation was found between these quantities in the deeper survey CDFS (Tozzi et al. 2006). 
As mentioned above, the current interpretation of this effect
is that there is an increasing contribution of the reflection component, that at high redshift (and hence luminosity)
enters the observing band. 
The decrease of the normalization of the reflection component with increasing luminosity is not strong enough to compensate this effect.

Making the plot of Fig. 10 (left) in logarithmic scale instead of linear, it would produce a distribution similar to
that of the $\Gamma$ vs \lum\ plot (right). The two quantities clearly go together in a flux/counts limited sample, but there is also 
a selection effect here, for which, at a given flux/counts, sources with flatter spectra have higher $L_{2-10 keV}$ (right).
Binning in linear z puts together sources with $L_{2-10 keV}$ in 1.5-2 dex range (at z\ltsima2), thus diluting the effect.

\subsection{Column density dependencies}

Fig. 11 (left panel) shows the distribution of intrinsic absorption, as a function of source redshift. 
We note that there is a strong dependency of the minimum value of \nh\ (either upper limit or detection) in the sample, and the redshift.
As already noted in Tozzi et al. (2006), this effect is due to the fact that, with increasing redshift, the photoelectric absorption cut-off moves 
toward the lower limit of the observing band (i.e. 0.5 keV), and it becomes more difficult to measure small values of \nh.
This can introduce a bias in the distribution of the observed fraction of obscured sources with redshift (Gilli et al. 2010).
To illustrate this effect, the dashed (dotted) curve in the plot represents the upper limit obtained from the spectral fit of a 200 (1000) counts 
simulated spectrum with no intrinsic absorption
and photon index $\Gamma=1.9$, at any given redshift. 

Furthermore, there is a lack of sources with high column densities (\nh$>10^{22}$ cm$^{-2}$) at low redshift ($z < 0.5$). 
This can be explained by the fact that a low luminosity, low redshift source, with a high column density
would have a highly suppressed flux (the photoelectric absorption cut off is at few keV), and it would not pass our counts--based selection criterion.
There are just two exceptions: one being the source CID-1105, at z=0.219.
This source shows, in addition to the primary power law, that is completely suppressed below 2.5 keV, a second soft component, 
which contributes for $\sim25\%$ to the total number of counts (see Sec. 4.3).
The other exception is source CID-1678 at z=0.104, that has a much lower column density and
also shows a hint of a second, soft component below 1 keV, but less prominent than in CID-1105.

Fig. 11 (right panel) shows the distribution of intrinsic absorption \nh\ as a function of the 2-10 keV intrinsic luminosity, corrected for absorption. 
The dashed vertical line marks the adopted separation between quasars and Seyfert (L$_{2-10 keV}=10^{44}$ erg s$^{-1}$),
the long dashed horizontal line represents the \nh\ value used to separate obscured and unobscured sources (\nh$=10^{22}$ cm$^{-2}$),
while the dotted vertical line shows the luminosity cut we adopted to distinguish between AGN and star forming galaxies (L$_{2-10 keV}=10^{42}$ erg s$^{-1}$, see Sec. 3.3).

As noted in Sec. 4.4 for the distribution of \nh\ and L$_{2-10 keV}$ separately,
type-2 AGN have higher values of \nh\ and lower luminosities, so that the region 
of \nh$>10^{22}$ cm$^{-2}$ and L$_{2-10 keV}<10^{44}$ erg s$^{-1}$ is populated 
almost only by type-2 sources, while almost all the sources in the opposite region 
(\nh$<10^{22}$ cm$^{-2}$ and L$_{2-10 keV}>10^{44}$ erg s$^{-1}$) are type-1s.
As mentioned for the \nh\ vs. z distribution, sources at low luminosities and high \nh\ are missed in our 
counts--limited sample due to selection effects.

About $15\%$ (62 out of 390) of all the sources lie in the region of obscured luminous quasar 
(\nh$>10^{22}$ cm$^{-2}$ at 1$\sigma$ and L$_{2-10 keV}>10^{44}$ erg s$^{-1}$).
Such high values of the fraction of X-ray selected type 2 quasar have been found in deeper \chandra\ and \xmm\ surveys 
(Mateos et al. 2005, Tozzi et al. 2006), and are compatible
with the predictions of X-ray background synthesis models 
($\sim12\%$ in the flux range of the CCBS, Gilli et al. 2007).
We underline here the importance of the  100\% completeness in redshift of CCBS:
typically not identified sources have faint optical counterparts, 
and these are usually obscured sources at high redshift-high L$_{2-10 keV}$ (Fiore et al. 2003). 
Most of the sources in the type-2 quasar region are classified as galaxies at high redshift, 
through their SED fitting. The excellent quality of COSMOS photometric redshifts
is indeed one of the unique features of COSMOS as an omni-wavelength survey, allowing to efficiently identify this class of elusive sources.

Twenty sources classified as type-1 AGN are in the region of type-2 quasar (out of 62). 
These are very interesting sources, given that they are natural candidates to be Broad Absorption Lines (BAL) quasars.
BAL quasars show broad, blue shifted absorption lines in the optical/UV spectra (Turnshek et al. 1980; Weymann et al. 1991).
They also show strong obscuration in X-rays, typically from a 'warm' absorber (Reynolds 1997; Piconcelli et al. 2005)
and/or in the form of blue shifted absorption lines of highly ionized gas (Fe XXV, Fe XXVI)
(Chartas et al. 2002; Markowitz et al. 2006; Tombesi et al. 2010; Lanzuisi et al. 2012).
Of these 20 BAL QSO candidates, 14 have an optical spectrum available, and almost all (13/14) of them lie at redshift 
$z>1.5$, i.e. it should be possible to observe such UV features (e.g. C IV at $\lambda1549$\AA)
from the ground in their optical spectra.
Indeed, to a visual inspection of their optical spectra, at least 50\% of them shows absorption signatures 
at the expected locations of ionized metals typically observed in BAL quasars, such as 
Mg II, Al III, Si IV, C IV, N V and O VI.
The proper analysis of the optical spectra of these BAL candidates is ongoing and will be the subject of a dedicated paper.


\subsection{X/O ratio}

\begin{figure}
\begin{center}
\includegraphics[width=8cm,height=8cm]{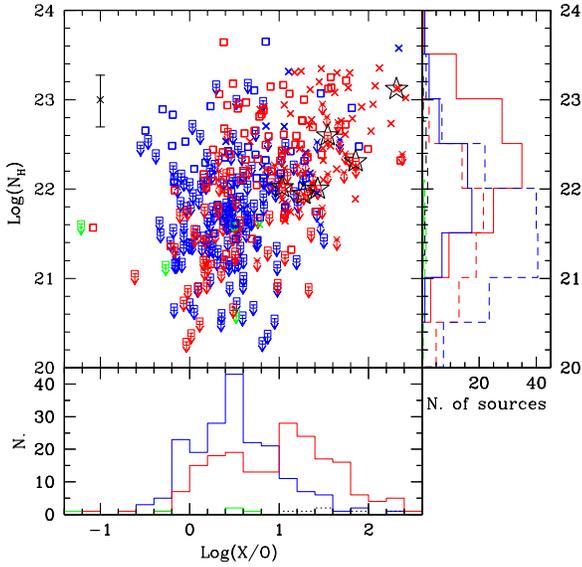}
\caption{ Distribution of X/O as a function of \nh, for the 390 sources in the CCBS. 
The bottom and right projections show the distribution of X/O and \nh\ respectively for each class of sources. Symbols and colors as in previous figures.}
\end{center}
\label{xo}
\end{figure}

We studied the distribution of the X/O ratio,
defined as in Civano et al. (2012), i.e. $X/O=log(F_X/F_{Opt})$\footnote{We used the 2-10 keV band and the $i$ optical magnitude.}.
It is well known that the X/O ratio is a z dependent quantity for obscured AGN, 
given that optical and X-ray bands have a k-correction going in opposite directions:
the k-correction is negative in the optical band 
and positive for the X-rays (Comastri et al. 2003; Fiore et al. 2003; Brusa et al. 2010).
As a result, obscured sources at high redshift have
higher X/O. On the other hand, unobscured sources have similar k-corrections
in the two bands, and the distribution in X/O is not correlated
with the redshift (see Fig. 14 of Paper III).

The X/O ratio is known to be a proxy for the absorbing column density, at least in the Compton thin regime (Fiore et al. 2003).
However, the actual correlation between \nh\ and X/O has not been demonstrated yet directly due to the lack of statistics in previous samples.
Fig. 12 shows the distribution of X/O as a function of \nh, for the 390 sources in the CCBS. 
The right and lower projections show the distribution of \nh\ and X/O, respectively, for each class of sources.
For type-1 AGN the distribution of X/O peaks at X/O $\simeq0.5$ and the majority 
of these sources are distributed between 0 and 1, while the
observed range for optically and soft X-ray selected AGN is typically more extended toward lower values 
(i.e.  $-1 <$ X/O $< 1$; Zamorani et al. 1999, Lehmann et al. 2001). 
This can be explained considering that we are sampling the bright end of the F$_{2-10 keV}$ distribution of AGN in C-COSMOS,
i.e. we are selecting the X-ray brightest sources. 

Type-2 AGN  peak at X/O $>1$, reflecting the fact that these sources are on average more obscured in the optical band, although a tail to low X/O and 
low \nh~is present. The median value for type-2s in our sample is $\langle X/O \rangle=0.96$, with dispersion of $\sigma=0.64$.
For comparison, the median value and dispersion for the sample of AGNs presented in Civano et al. 2012, combining 
\chandra-COSMOS and \xmm-COSMOS X-ray catalogs are $\langle X/O \rangle=1.02_{-0.72}^{+0.59}$ and $\langle X/O \rangle=0.82_{-0.74}^{+0.66}$
for type-2s and galaxy SEDs respectively. 

Among these sources there is a large fraction of sources with optical/IR properties of passive galaxies, that are classified
as AGN thanks to the multi-wavelength coverage and SED fitting together with X-ray informations.
Indeed, the typical \nh\ for these sources is in the range $10^{22}-10^{23}$ cm$^{-2}$
and, if converted into optical extinction $A_V$ through the standard Galactic dust-to-gas ratio ($A_V =$ \nh$ / (1.8\times10^{21})$, e.g. Predehl \& Schmitt 1995), 
translates into an extinction $A_V\simgt5$, large enough to completely obscure the AGN contribution in the optical bands. 

The distribution of X/O vs. \nh\ for type-2 AGN closely resembles the one found between \nh\ and 2-10 keV luminosity.
This is due to the fact that mildly obscured AGNs (up to few$\times10^{22}$ cm$^{-2}$)  
have their nuclear optical/UV light strongly attenuated by dust and, therefore, 
the X/O ratio represents the ratio between the X-ray flux from the AGN and the host galaxy starlight, 
which spans a smaller luminosity range (Fiore et al. 2003).

Almost all type-1 AGN with \nh$>10^{22}$ cm$^{-2}$ (i.e. BAL quasar candidates),
have an X/O ratio in the range 0-1, typically lower than the population of type-2 obscured sources.
This is in agreement with the idea that these sources are obscured by outflowing gas in the very central regions,
blocking the X-rays coming from the inner regions of the system, but not the optical emission coming from 
much larger regions.

Nearly 70\% of all the sources with X/0$>1$ have \nh$>10^{22}$ cm$^{-2}$,
thus confirming the idea that the X/O ratio is a proxy for the obscuring column density, 
and selecting sources above this threshold is an efficient tool for the creation of samples of obscured AGN (Fiore 2003; Comastri \& Fiore 2004).
However, there is a 30\% of contamination by sources with high Log(X/O) and low X-ray obscuration, as found in Tozzi et al. (2006).

\section{Summary}

We have performed the spectral analysis of the 390 brightest extragalactic sources in the \chandra-COSMOS field with more 
than 70 net counts in the 0.5-7 keV band.
The sample has a 99\% completeness in optical identification, $\sim75\%$ of the sources 
have a spectroscopic redshift, while the remaining $\sim25\%$ have a photometric redshift.

The spectral fit was performed using a non binned minimization technique and assuming as underlying model a simple power law, modified by 
intrinsic absorption at the redshift of the source, plus galactic absorption fixed to the 
value in the direction of the field. Simultaneously a global model for the background was fitted to the local background spectra, leaving 2 free parameters
to account for local variations.

The main observational results can be summarized as follows:
\begin{itemize}
\item In appendix A we show that our spectral fit procedure, based on the use of Cstat statistic and simultaneous total and background fit,
proved to be accurate and reliable:
the spectral parameters obtained are stable, regardless of the combination of statistic and number of counts per bin we chose,
while for the errors we proved that the Cstat applied to binned spectra (with 5 counts per bin) gives more constrained results
(i.e. relative errors $<30\%$ for most sources) with respect to a $\chi^2$ analysis.
The Cstat applied to non binned spectra
tends to underestimate the errors.

\item The sample is divided almost equally in type-1 (49.7\%) and type-2 AGN (48.7\%)
plus few passive galaxies at low z.
However 70\% of type-2 AGN has a spectral or photometric
classification of passive/starforming galaxy, or with hybrid or ambiguous classification,
and an 'a posteriori' selection criterion, based on the X-ray luminosity, was required 
to identify them as sources hosting a hidden AGN.
Indeed, at the flux level of our sample almost all the 
sources classified as non active turned out to be obscured AGN.

\item The average photon index of type-1 AGN is $\langle  \Gamma \rangle =1.89 \pm 0.02$
with intrinsic dispersion $\sigma_{int}=0.11$, while for type-2 AGN is $\langle  \Gamma \rangle =1.76 \pm 0.03$
and $\sigma_{int}=0.13$. The two distributions have a high probability to be intrinsically different
(KS test probability $4.04\times10^{-5}$).
With the current data it is not possible to determine if this is an intrinsic difference or
it is due to unaccounted extra components in the spectra.

\item The distribution of column density for type-1 AGN is dominated by upper limits (75\%), anyway 
$\simeq$17\% of them results obscured from the X-ray spectral analysis. 
For type-2 AGN the distribution is broader, with a peak around $10^{22}$ cm$^{-2}$
and 60\% of the sources with a detection of the value of the column density.
Only $\sim50\%$ of type-2 can be considered obscured. 
This is a low value compared with previous results.
Possible explanations can be a significant contribution from starbursts emission in the soft band,
or misclassification of faint type-1 with  strong optical/IR contamination from host galaxy light.

\item No obvious CT source is present in the CCBS, even 
if a sample of 6 CT candidates has been selected on the basis of their very flat X-ray spectra.
For only 2 of these sources, an emission Fe line is detected, at energies slightly different, but still consistent with
the expected value.

\item The intrinsic 2-10 keV luminosity, corrected for absorption, of type-1 AGN peaks at 
$\simeq10^{44}$ erg s$^{-1}$ and 63\% of these sources are in the quasar regime. 
The majority ($\sim70\%$) of type-2 AGN instead are concentrated below this value and are therefore mostly 
Seyfert galaxies.
This difference is clearly due to the combination of an X-ray flux limited sample and the different redshift
distributions, with type-1 AGN having typically higher redshift.

\item No correlation has been found between the photon index and the redshift for our sources, while a weak 
anti-correlation is observed in X-ray luminosity. The correlation (with slope $-0.09\pm0.06$)
is weaker of what has been found in previous work on \chandra\ and \xmm\ counterpart of bright SDSS quasars,
while consistent with no correlation as found in the CDFS.

\item  About $15\%$ of the sources lie in the region of type-2 QSOs, 
a value in agreement with deeper X-ray surveys and with population synthesis models. 
This high fraction was made possible by the almost 100\% completeness of our sample and the accuracy of photometric redshifts: a large fraction of
these QSO2 are indeed high redshift galaxies with clean counterpart only in near and mid IR.
20 sources are type-1 BAL QSO candidates. Of them 14 have an optical spectrum and at least 50\% of them indeed show indication
of absorption lines in their optical spectra.

\item The distribution of X/O for type-1 AGN peaks at X/O $\simeq0.5$ and the vast majority of these 
sources are distributed between 0 and 1. 
Type-2 AGN peak at X/O $\simgt1$ and 50\% of them show $1 <$ X/O $< 2$, an extreme value for AGN.
Furthermore, nearly 70\% of all the sources with X/O $>1$ are obscured, thus confirming that selections 
based on high X/O ratio are efficient in finding samples of obscured AGN. 

\item In appendix B we show that there is a small systematic difference in the spectral parameters
obtained from \chandra\ and from \xmm\ (280 \xmm\ counterparts of the CCBS sources).
The \xmm\ spectra tend to be fitted with softer power laws (up to 20\% difference) and lower obscuration (5\% difference in
the fraction of obscured sources) with respect to \chandra\ spectra. 

\end{itemize}

The CCBS represents just the tip of the iceberg of the complete C-COSMOS catalog,
and a large number of interesting sources such as heavily obscured AGN and starbursts galaxies 
are located below our selection threshold. 
The use of our well tested procedure on this larger sample will give important insights on the 
co-evolution of accretion and star formation.

Finally, the \chandra\ COSMOS-Legacy large program has been recently approved (Cycle 14), 
aimed to extend the \chandra\ coverage of the COSMOS area,
taking $\sim1.45$ deg$^2$ to 160 ks depth and covering the remaining 0.75 deg$^2$ field to 100-50 ks depth (for a total of 2.8 Msec).
It is expected to produce a final sample of bright sources almost 3 times larger than the current one ($\sim1100$ sources)
allowing for a much more accurate determination of the intrinsic distribution  of the X-ray spectral parameters
in the sample.

\section*{Acknowledgments}
We thank the anonymous referee for providing constructive comments and help in improving the contents of this paper.
We acknowledge financial contribution from the agreement ASI-INAF I/009/10/0.
This work was supported in part by
NASA Chandra grant number GO7-8136A (M.E., F.C.),
the Blancheflor Boncompagni Ludovisi foundation (F.C.), and
the Smithsonian Scholarly Studies (F.C.).
This work is based on observations
made with the \chandra\ X-ray satellite.
Also based on observations obtained with \xmm\, an ESA science mission
with instruments and contributions directly funded by
ESA Member States and NASA.
This research has made use of data and/or software provided by the 
High Energy Astrophysics Science Archive Research Center (HEASARC),
which is a service of the Astrophysics Science Division at NASA/GSFC 
and the High Energy Astrophysics Division of the Smithsonian Astrophysical Observatory.

\begin{appendix}

\section{Cstat and $\chi^2$}

\begin{figure}[h]
\begin{center}
\includegraphics[width=6cm,height=6cm]{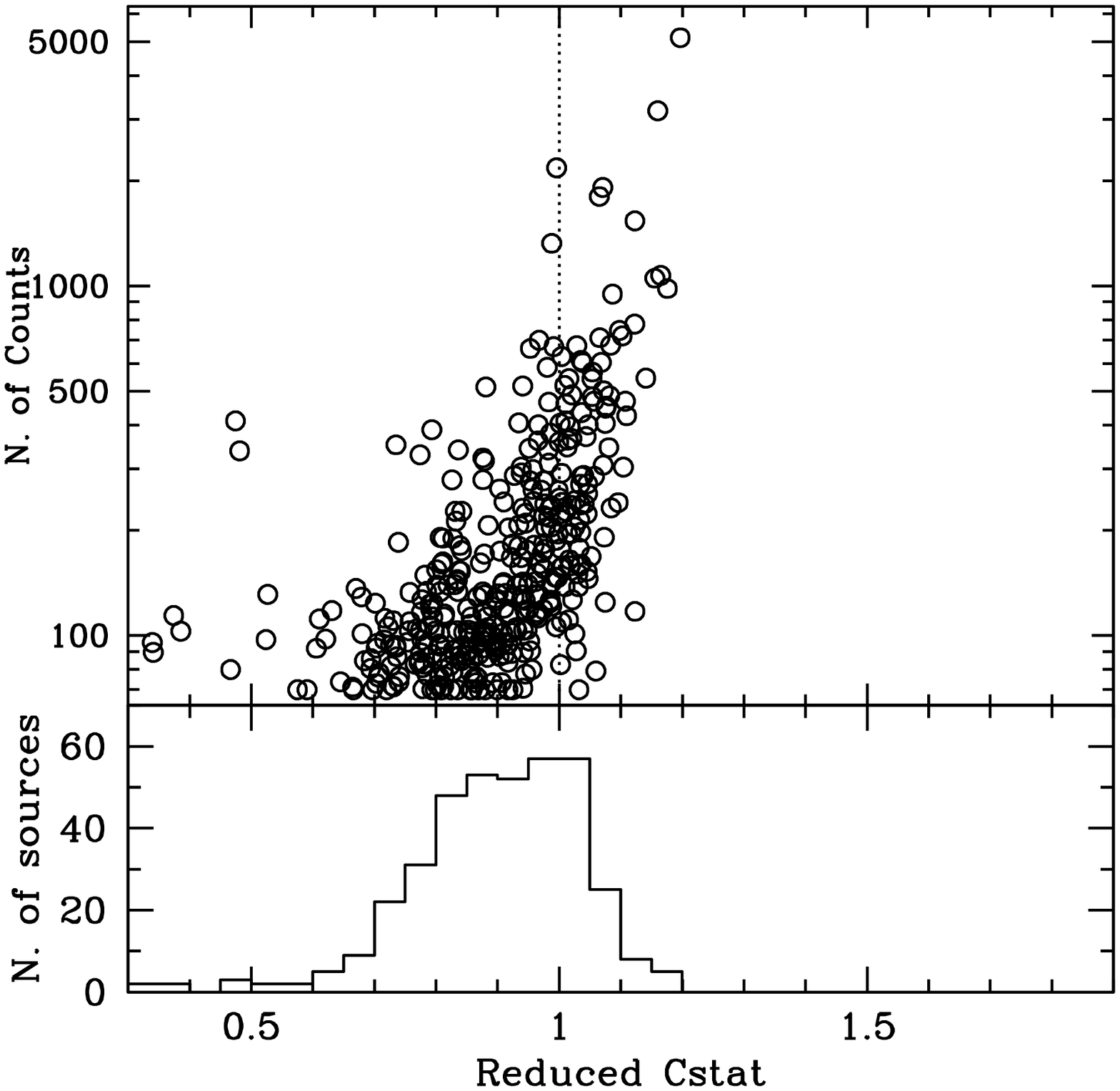}\hspace{1.5cm}
\includegraphics[width=6cm,height=6cm]{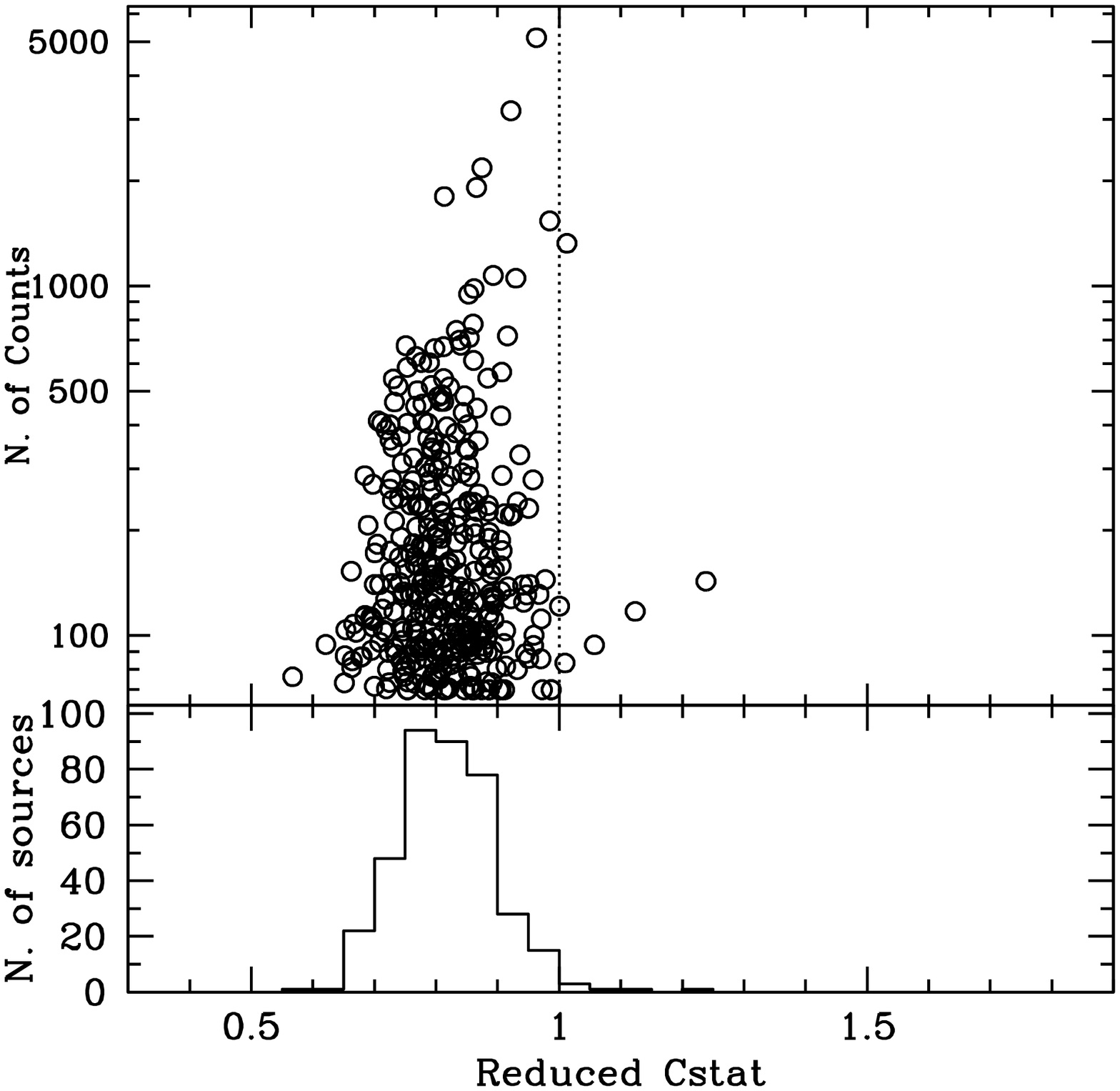}
\caption{Distribution of Cstat$_\nu$ as a function of counts for the not binned case (top) 
and the 1 counts per binned case (bottom).}
\end{center}
\label{cstat_app}
\end{figure}

\begin{figure}
\begin{center}
\includegraphics[width=5cm,height=5cm]{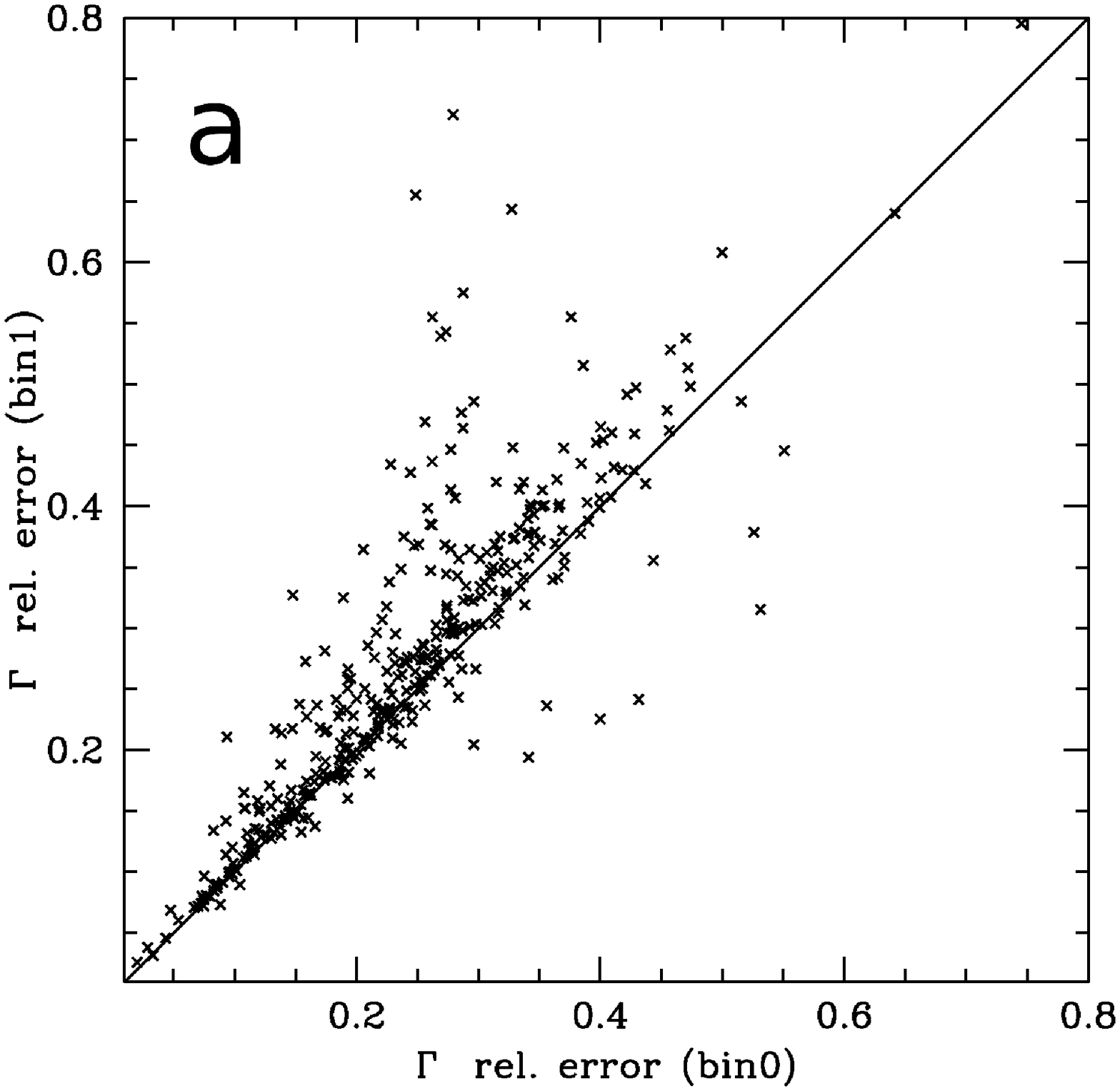}\hspace{0.1cm}
\includegraphics[width=5cm,height=5cm]{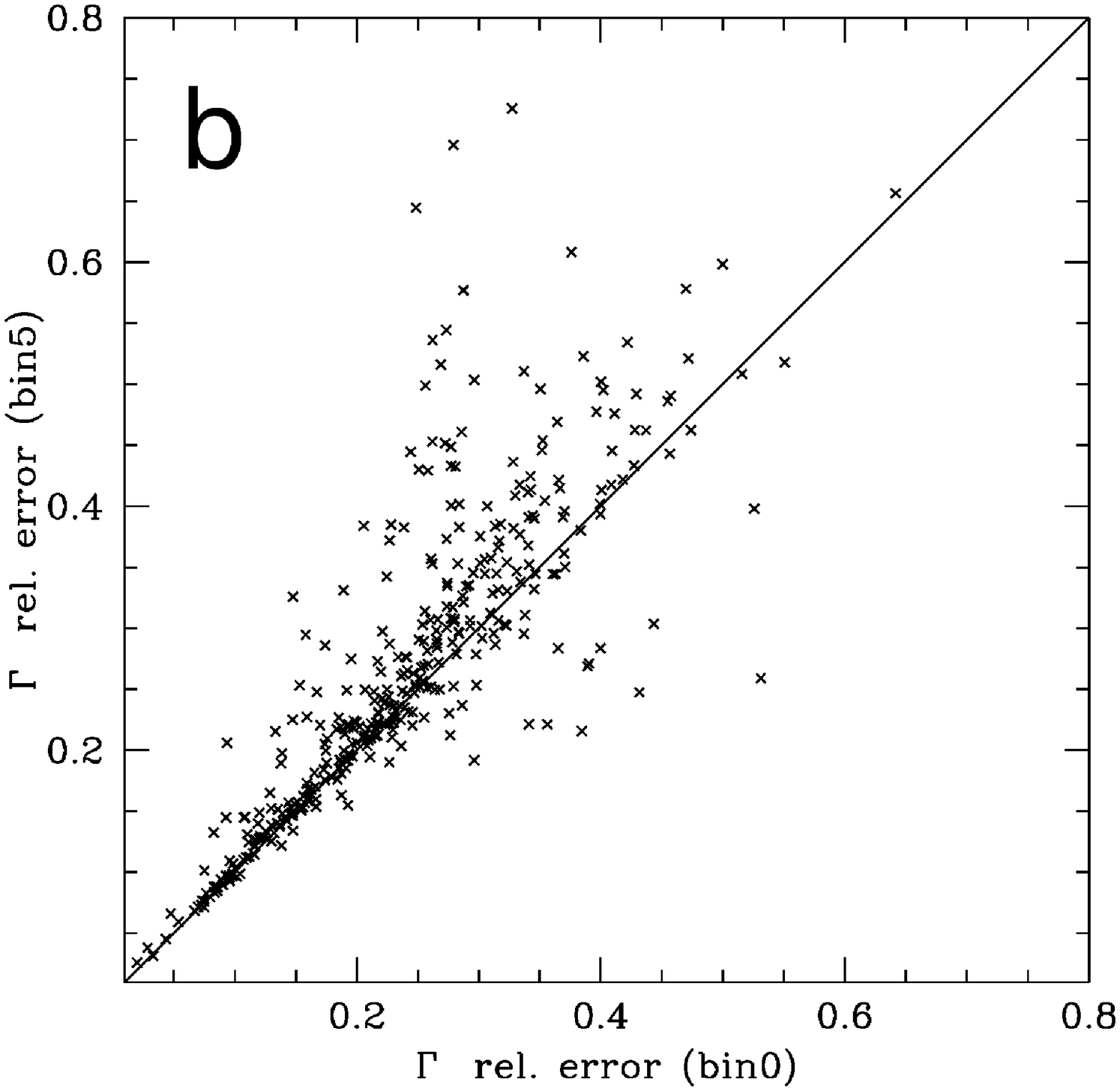}

\vspace{0.1cm}

\includegraphics[width=5cm,height=5cm]{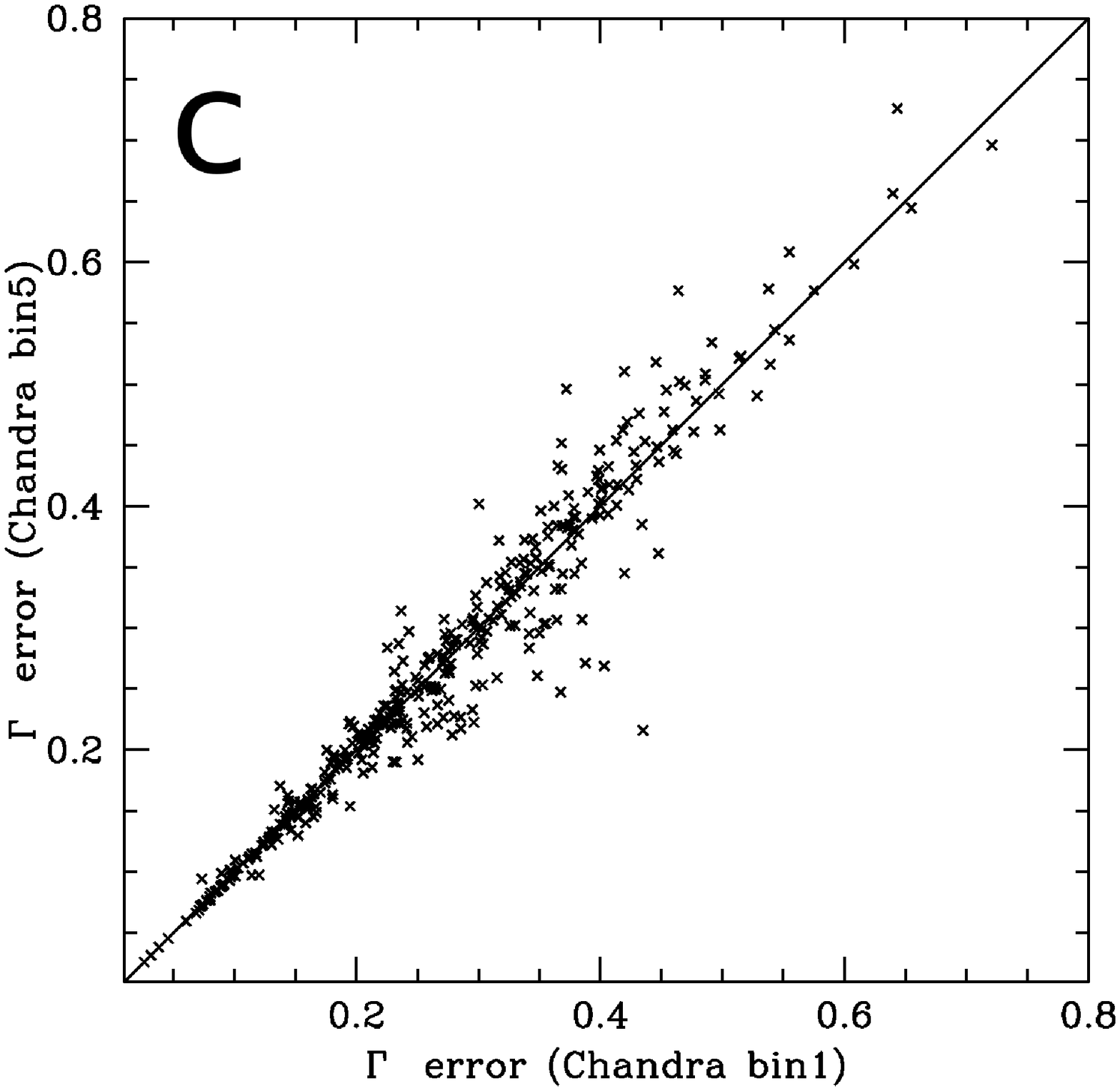}\hspace{0.1cm}
\includegraphics[width=5cm,height=5cm]{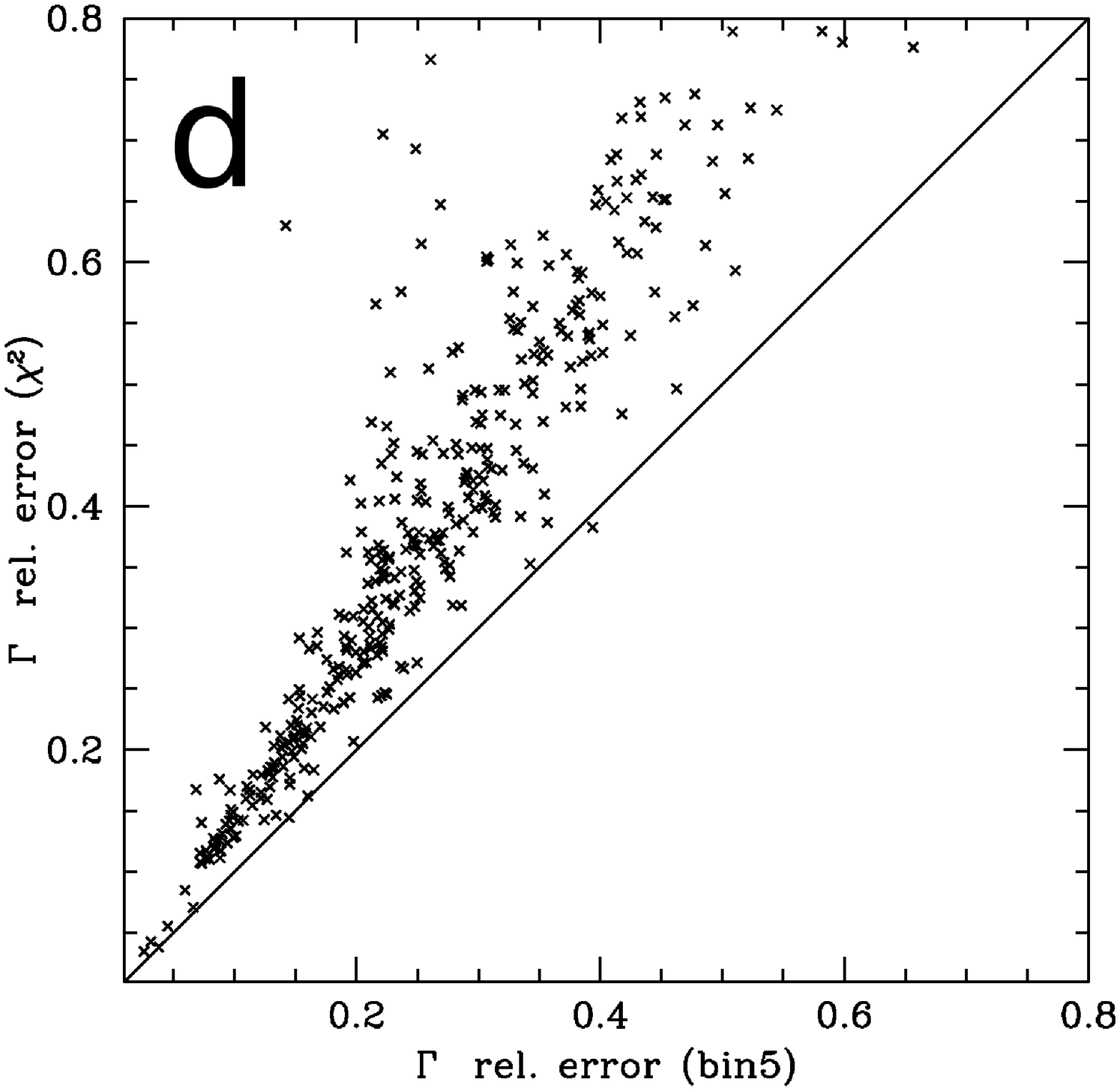}
\caption{Comparison between the relative errors on $\Gamma$
for different binning and statistics adopted: Cstat(bin0) vs. Cstat(bin1); 
Cstat(bin0) vs. Cstat(bin5); Cstat(bin1) vs. Cstat(bin5) and Cstat(bin5) vs. $\chi^2$(bin15), from top to right).}
\end{center}
\label{ergam}
\end{figure}

We performed a series of tests to verify the behavior of the Cstat statistic in the range of counts available 
for our sample, from 70 to few thousands net counts.
We verified that the Cstat$_\nu$, applied to unbinned data, is strongly dependent on the number of counts,
moving to $\sim1$ for sources with few thousands counts, to $\sim0.7-0.8$ for sources with $70-100$ counts (Fig. A.1 top panel).
This happens because of the high number of empty channels for the sources with few tens of counts,
which contribute to the total number of DOF, but not to the amount of Cstat.
On the other hand, applying a binning of 1 counts per bin removes the dependency of Cstat$_\nu$ from 
the number of counts, because the empty channels are avoided.
However, the distribution of Cstat$_\nu$ is not centred at $\sim1$ as expected (and as for the 5 binning case)
but instead  at lower values ($\sim0.85$, Fig. A.1 bottom panel). 

Interestingly, the resulting spectral parameters obtained in the three ways are almost indistinguishable. 
Also  the associated errors are comparable, even if, when no counts binning is applied,
there is a population of sources for which the errors are typically smaller, with respect to both fits with 1 and 5 counts per bin.
Fig. A.2 (panel a) shows the comparison between the relative errors on $\Gamma$ measured 
from Cstat with no binning and from Cstat with 1 counts per bin.
Panel b shows the comparison between Cstat with no binning and Cstat with 5 counts per bin.
Not surprisingly, the average difference between errors increases with higher values of relative errors, i.e. lower counts available in the spectra.
Panel c shows the comparison between Cstat with 1 counts per bin and Cstat with 5 counts per bin, that are comparable.

To estimate the expected errors on the spectral parameters, we simulated 1000 spectra in four counts regime, i.e. 100, 200 500 and 1000 net counts, respectively.
The input model has fixed values of $\Gamma=1.9$, \nh$=1\times10^{22}$ cm$^{-2}$ and $z=1.0$ (i.e. close to the average value for our sample).
We then performed our analysis with the Cstat and the three different binning (bin0, bin1 and bin5).
The $1\sigma$ dispersions on $\Gamma$ are comparable in the three cases, and of the order of 0.25, 0.19, 0.15 and 0.09,
that translate into relative errors of 13,10,8 and 5\% respectively, thus fully in agreement with the distribution of relative error
show in Fig. 5 (right panel).

 We also performed a fit applying the $\chi^2$ statistic for all the sources, 
after binning the spectra to 15 counts per bin.
The best fit spectral parameters are again very similar to those derived from the Cstat statistic (regardless of the number of counts per bin).
Instead, the average ratio between the relative errors obtained with Cstat (5 counts per bin) and with the $\chi^2$ 
is $Err_{Cstat}/Err_{Chi^2}=0.69$, with a small dispersion ($\sigma=0.11$), 
i.e. the Cstat gives error typically $\sim30\%$ smaller than $\chi^2$, for our sources (see also Tozzi et al. 2006), 
with increasing differences at decreasing number of counts,
enforcing our choice to use the Cstat instead of $\chi^2$ statistic.
In Fig. A.2 (d panel) is reported the comparison between the relative errors on $\Gamma$
computed from Cstat with 5 counts per bin and from $\chi^2$ with 15 counts per bin.

\section{Comparison with XMM-Newton COSMOS data} 

The COSMOS field has been observed in X-rays by \xmm\ between 2003 and 2006.
The \xmm\ observation covers a region of $1.4\times1.4$ deg$^2$, 
to a sensitivity limit of $0.7\times10^{-15}$ erg cm$^{-2}$ s$^{-1}$ in the 0.5-2 keV band
(Hasinger et al. 2007).
Details on the point-source detection method for the XMM-COSMOS survey are presented in Cappelluti et al. (2007),
while details on extraction of the X-ray spectral counts for the sources in the XMM-COSMOS catalog, are described in Mainieri et al. (2007).
As for C-COSMOS sources, the spectral counts were extracted separately from each observations and merged together, weighting the 
response matrices ARF and RMF for the contribution of each spectrum.
We applied the same cut in number of counts (70 net counts in 0.5-7 keV) 
to collect a sample of 280 \xmm\ counterparts of the CCBS sources.

The spectral analysis was performed in the same way described in Sec 4, i.e. 
using the Cstat minimization technique, applying a 5 counts binning and simultaneously fitting the background.
The energy band taken into account was limited to the 0.5-7 keV band, equal to the one used for the fit of \chandra~ data, 
even if the nominal sensitivity of \xmm\ reaches $\sim10$ keV, in order to minimize systematic effects due to the different bandpass.

\subsection{Photon index}

\begin{figure*}
\begin{center}
\includegraphics[width=7cm,height=7cm]{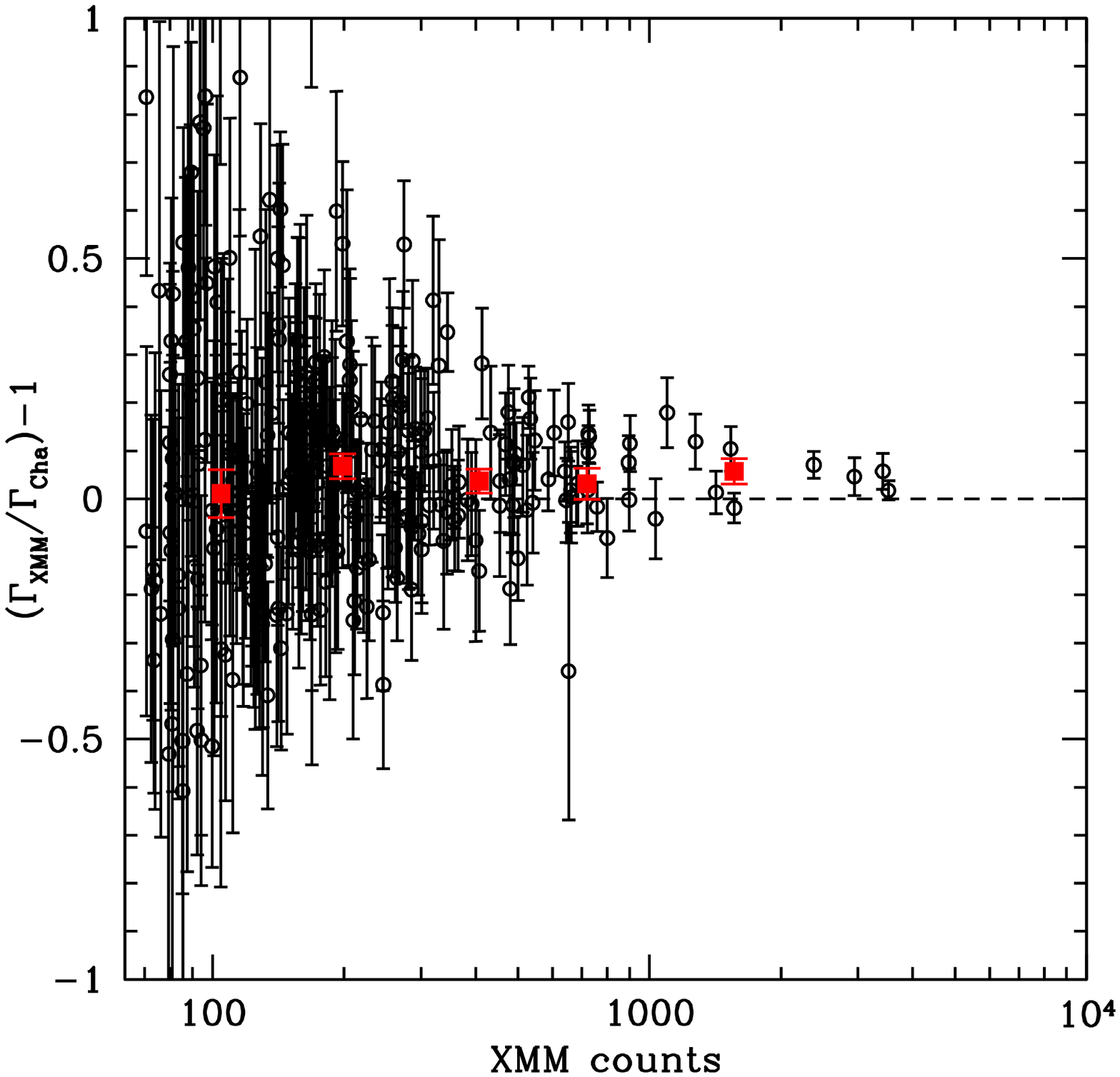}\hspace{1.5cm}
\includegraphics[width=7cm,height=7cm]{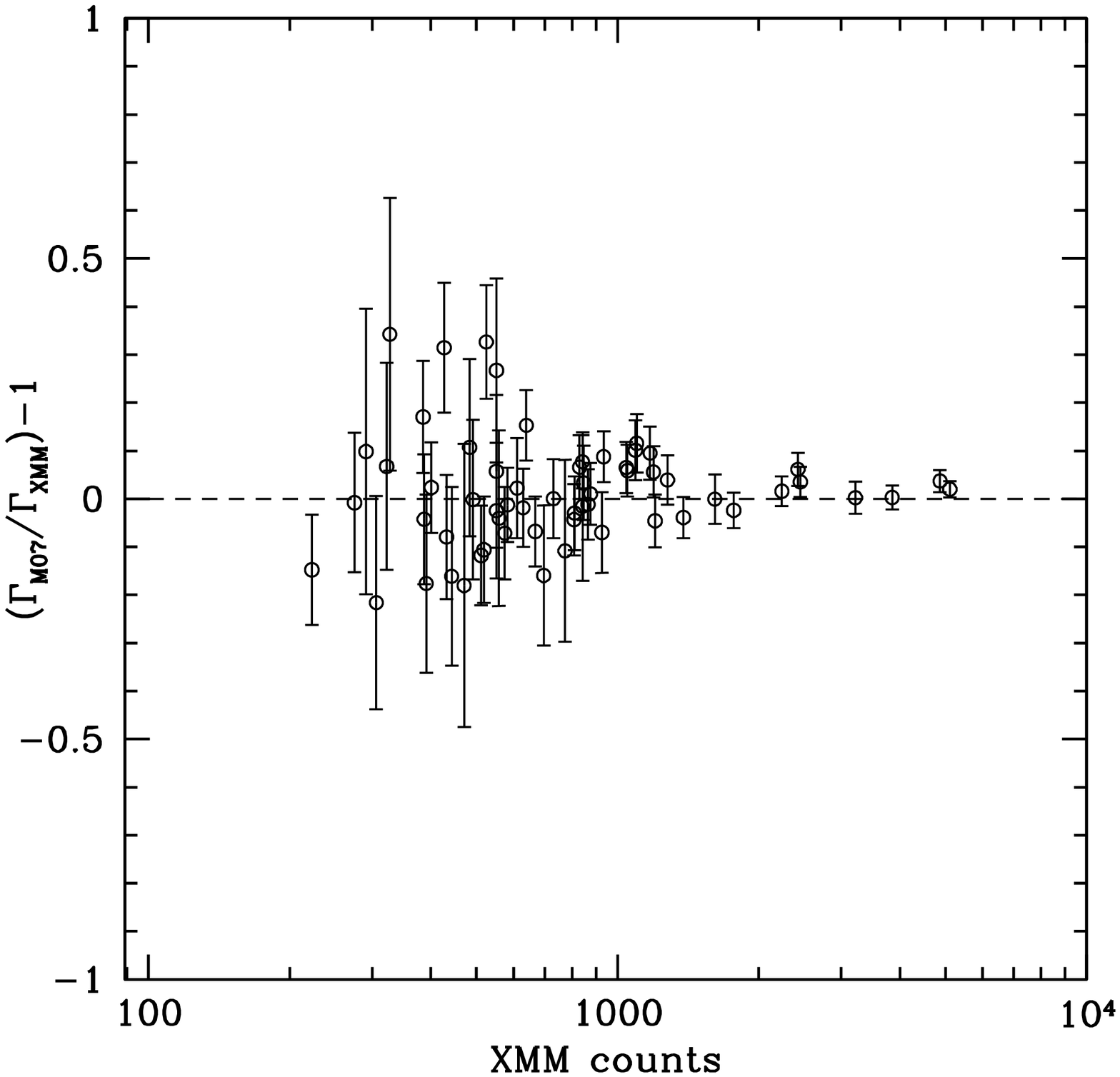}
\caption{{\it Left panel:} Normalized difference in photon index (($\Gamma_{XMM}/\Gamma_{Chandra})-1$) 
as a function of the number of \xmm\ counts for all the 280 sources with \xmm\ counterpart showing more than 70 counts. 
The red squares represent the median of the distribution of $\Delta\Gamma/\Gamma$ computed in bins of counts (70-150; 150-300; 300-600;
600-1000 and $>$1000). The vertical bars show the standard error on the median for each bin. 
{\it Right panel:} Normalized difference between the photon index obtained from our analysis of \xmm\ data and the one 
independently obtained in Mainieri et al. 2007, for the 58 sources in common.}
\end{center}
\label{gamma}
\end{figure*}

Fig. B.1 (left panel) shows the normalized difference in photon index 
(($\Gamma_{XMM}/\Gamma_{Chandra})-1$, i.e. $\Delta\Gamma/\Gamma=0.1$
is  equivalent to a 10\% difference in $\Gamma$) as a function of the number of \xmm\ counts for all the 280 \xmm\ counterparts of the CCBS sources. 
The red squares represent the median of the distribution of $\Delta\Gamma/\Gamma$, computed in bins of counts (70-150; 150-300; 300-600;
600-1000 and $>$1000). We note that there is a general good agreement between the two values,
but for bright sources (i.e. with more than $\sim1000$ counts) the distribution is asymmetric and the difference tend to be always positive,
i.e. the \xmm\ photon index is typically higher than the \chandra\ one, with the difference spanning the range 0-20\%.
Furthermore for most of these sources the relative difference is not consistent with zero even taking in to account the error bars. 
This trend is present, although less significantly because of the larger errors, in the bins with a small number of counts.

Given this unexpected result, we decided to further investigate this issue. We collected the photon indices computed in Mainieri et al. (2007)
for a sample of 135 AGN in XMM-COSMOS survey, showing more than 100 total counts and having a spectroscopic identification. 
The spectral fit of these sources was performed in a completely different way, using the $\chi^2$ minimization technique, 
and performing background subtraction and counts binning. 58 of these sources are in common with our sample. 
Fig. B.1 (right panel) shows the relative difference between the photon index obtained from our analysis of \xmm\ data 
and the one independently obtained in Mainineri et al. (2007), as a function of \xmm\ counts for these sources. 
These values are in very good agreement, at all counts level, and in any case consistent within the error bars for almost all the sources.
This analysis suggests that we are observing an unexpected systematic difference in the determination of the photon index of the order of 
$\sim10\%$ from \chandra~ and \xmm, with \xmm\ spectra being softer than \chandra\ ones. 

In our analysis of \chandra\ spectra, we used as calibration files CALDB 4.1.2 for the extraction of \chandra\ spectra and matrices.
It is known that a major correction in the calibration files occurred in Dec. 2010, due to 
the contamination model, which under-predicted the attenuation by the Ca-K edge, starting from 2009\footnote{http://cxc.cfa.harvard.edu/ciao/why/acisqecontam.html}.
A similar, but less severe, correction occurred in Dec. 2009, regarding observations taken from 2006 (C-COSMOS observations were taken mostly during 2007).
As a check we extracted, with the newest calibration files available (CALDB 4.3), the spectra and matrices of the 20 brightest sources ($>500$ counts) in the sample.
Then we performed again the same spectral analysis on the new spectra, to verify if this new version of the calibrations
has some effect on the determination of spectral parameters.
The results obtained with this new calibration are completely consistent with the ones described above.

\subsection{Column density}

\begin{figure*}
\begin{center}
\includegraphics[width=7cm,height=7cm]{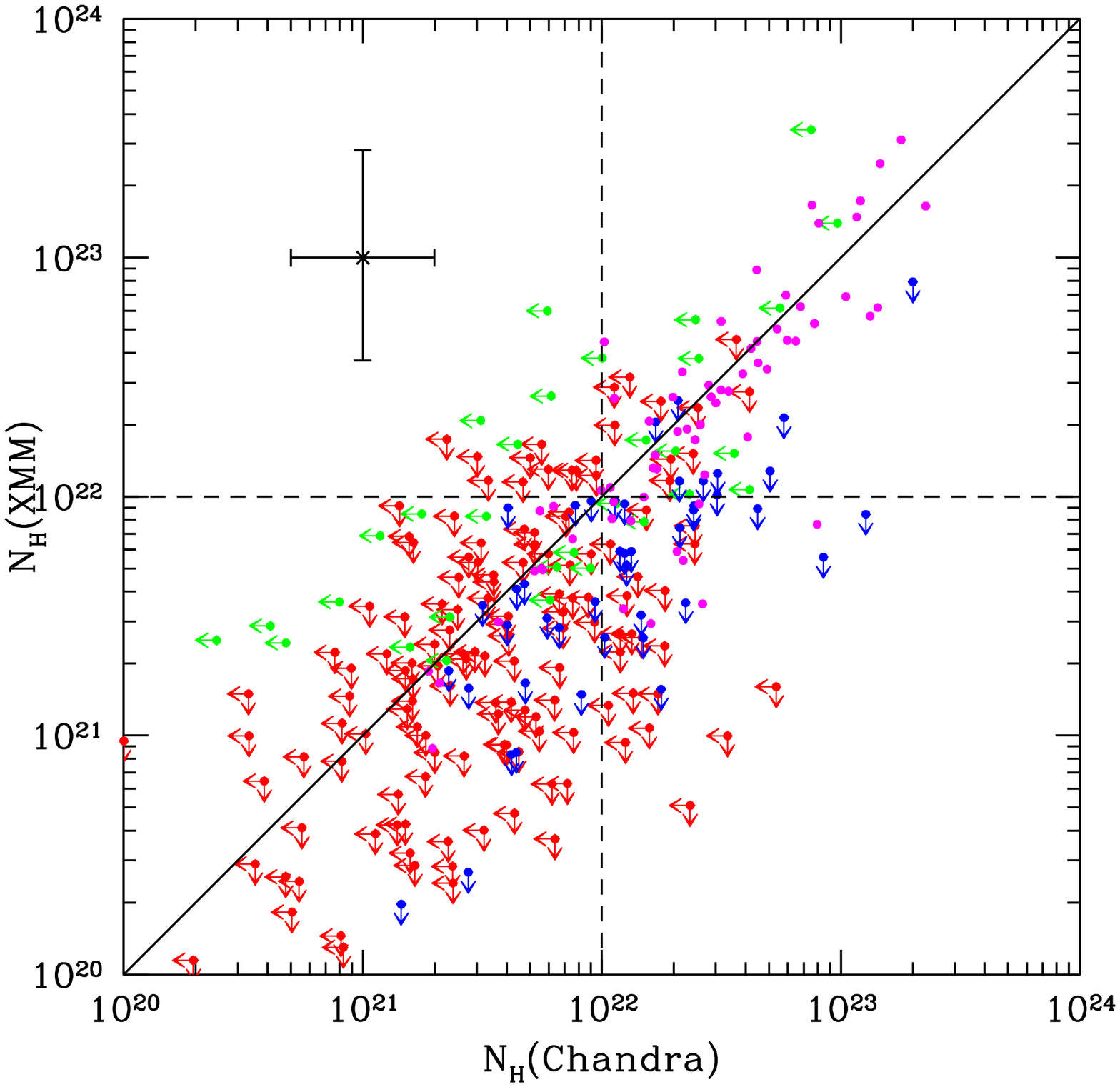}\hspace{1.5cm}
\includegraphics[width=7cm,height=7cm]{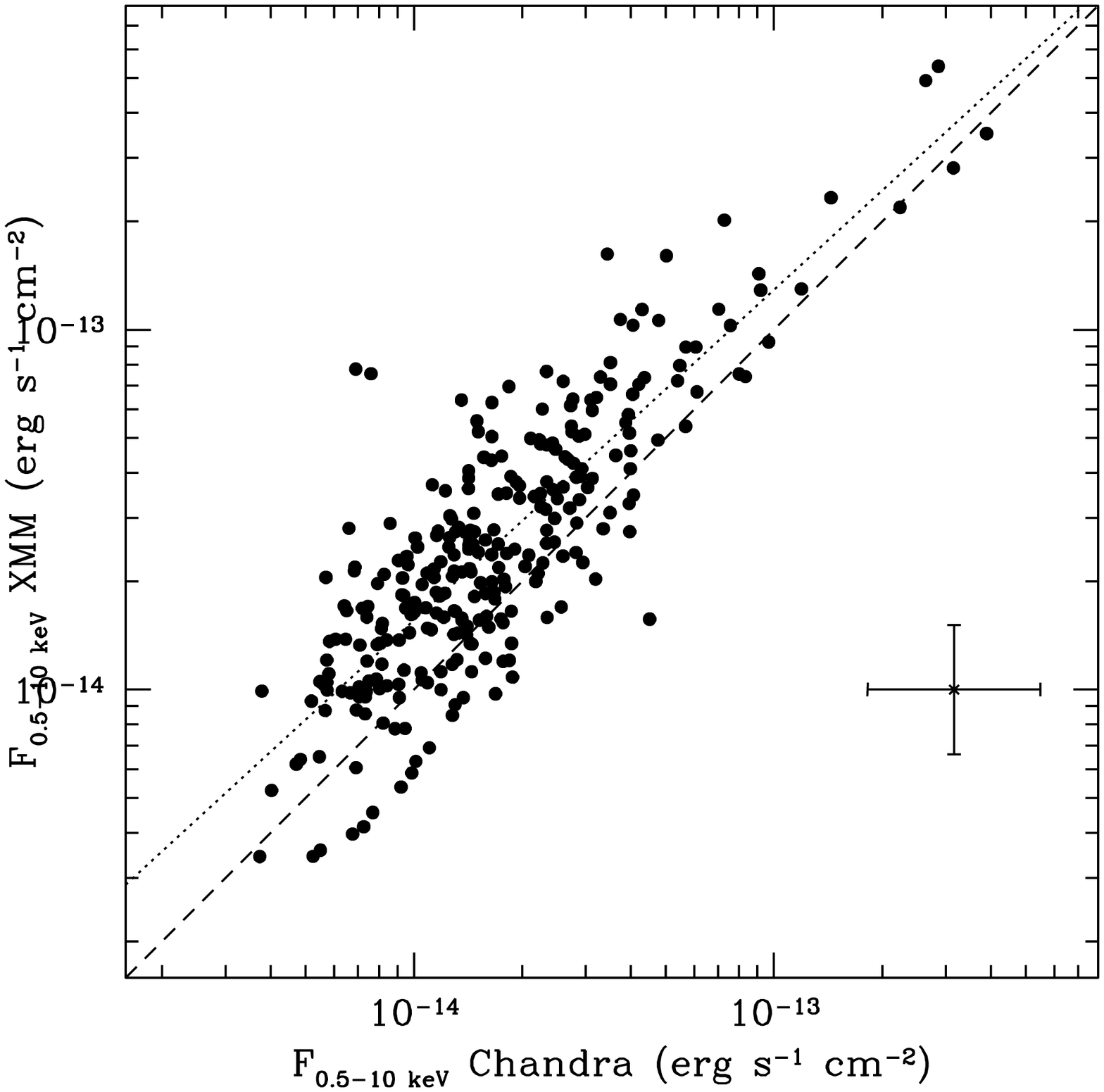}
\caption{{\it Left panel:} Comparison between the columns density obtained from the fit of \chandra\ data, 
and the \nh\ obtained from \xmm\ for the 280 counterparts of the CCBS. 
Different colors represent different combinations of detection and upper limits in the measured value of \nh\ (see text).
The thick line represents the 1:1 relation, the dashed lines mark the separation between 
obscured and unobscured sources, in both axes. 
{\it Right panel:} Comparison between the 0.5-10 keV band fluxes obtained from the two sets of data for the sample of 280 sources in 
common in the two surveys. The dashed line represents the 1:1 relation, the dotted line is the linear regression obtained for the data.}
\end{center}
\label{confronto}
\end{figure*}

Fig. B.2 (left panel) shows the comparison between the columns density obtained from the fit of \chandra\ data and the 
\nh\ obtained from \xmm\ for the sample of 280 sources in the CCBS with \xmm\ counterpart.
Magenta points represent \nh\ detections from both data sets,
green (blue) points represent \chandra\ (\xmm) upper limits and \xmm\ (\chandra) detection of the \nh, and red points represent 
upper limits on \nh\ from both datasets
The thick line represents the 1:1 relation, and the dotted lines mark the threshold assumed to distinguish between 
X-ray obscured and unobscured sources (\nh=10$^{22}$ cm$^{-2}$).
For sources with a detection on \nh\ from both satellites, there is a good agreement in the values of \nh\ 
along all the range of values. 
We note that there is a small population (6 sources) that occupies the lower right quadrant 
(i.e. \chandra\ measures a column density \nh$>10^{22}$ cm$^{-2}$ while the value measured by \xmm\ is lower)
while the opposite region (upper left) is empty. 
Similarly, there are slightly more sources which result 
obscured from \chandra\ and for which \xmm\ measures an upper limit (15), than the opposite (11 sources).
This translates in a small difference on the global fraction of obscured sources in the sample, that drops from $\sim40\%$ to $\sim30\%$ 
considering \chandra\ or \xmm\ results respectively.
This could in principle partly account for the mismatch historically observed between data
from the Msec-deep \chandra\ fields, namely CDF-S and CDF-N and results from \xmm\ surveys, 
at fluxes where these samples overlap (e.g. Fig. 16 of Gilli et al. 2007).

\subsection{Flux}

Fig. B.2 (right panel) shows the comparison between the 0.5-10 keV band fluxes obtained from the two data sets 
for the sample of 280 sources in common.
The dashed line represents the 1:1 relation, while the dotted line has been obtained as the linear regression of the two set of values. 
The average off-set increase with decreasing fluxes, being $\sim20\%$ at F$_{0.5-10}=10^{-13}$ erg s$^{-1}$ cm$^{-2}$
and increasing to $\sim40\%$ at  F$_{0.5-10}=10^{-14}$ erg s$^{-1}$ cm$^{-2}$.
However, for single sources, the values for the two data sets are consistent within the error bars, for most of the sources with 
F$_{0.5-10}<7-8\times10^{-14}$ erg s$^{-1}$ cm$^{-2}$ while sources with brighter fluxes have much smaller error bars,
except for source CID-1678, that is at the very edge of the \chandra\ pointings,
and tent to be not consistent with the 1:1 relation.
The general trend can be partly due to a selection effect: we start from the \chandra\ sample,
and then apply the counts cut to the \xmm\ catalog. Thus we tend to select select sources with a positive fluctuation in flux in the \xmm\ final sample.
Finally, a few \% of the sources shows a difference in the fluxes up to a factor 10, clearly indicating a real, intrinsic variability. 
However, the study of the X-ray variability, between the \chandra\ and \xmm\ data, and inside each set of data 
and its relation with the optical variability (see Salvato et al. 2012, Lanzuisi et al. 2012 in preparation) 
will be the subject of a dedicated paper.

\end{appendix}

\end{document}